\documentclass[aps,prx,superscriptaddress,twocolumn]{revtex4-2}
\usepackage{graphicx,color}
\usepackage{verbatim}
\usepackage{amssymb}   % for math
\usepackage{amsmath}
\usepackage{amsfonts}
\usepackage{mathdots}
\usepackage{hyperref}
\usepackage{epsfig}
\usepackage{hyperref}
\usepackage{xcolor}
\definecolor{myblue}{RGB}{46, 48,146}
\hypersetup{colorlinks=true,linkcolor=myblue,citecolor=myblue,urlcolor=myblue, linktocpage}
\usepackage{braket}
\usepackage{bm}
\usepackage{multirow}

\begin{document}
	\title{Gate-defined topological Josephson junctions in Bernal bilayer graphene }
	
	\author{Ying-Ming Xie}
	\affiliation{Department of Physics, Hong Kong University of Science and Technology, Clear Water Bay, Hong Kong, China} 	
	\affiliation{Department of Physics, California Institute of Technology, Pasadena, California 91125, USA}
	\affiliation{Institute for Quantum Information and Matter, California Institute of Technology, Pasadena CA 91125, USA}
	
	\author{\'Etienne Lantagne-Hurtubise}
	\affiliation{Department of Physics, California Institute of Technology, Pasadena, California 91125, USA}
	\affiliation{Institute for Quantum Information and Matter, California Institute of Technology, Pasadena CA 91125, USA}
	
	\author{Andrea F. Young}
	\affiliation{Department of Physics, University of California at Santa Barbara, Santa Barbara CA 93106, USA}
	
	\author{Stevan Nadj-Perge}
	\affiliation{T. J. Watson Laboratory of Applied Physics, California Institute of
		Technology, 1200 East California Boulevard, Pasadena, California 91125, USA}
	\affiliation{Institute for Quantum Information and Matter, California Institute of Technology, Pasadena CA 91125, USA}
	
	\author{Jason Alicea}
	\affiliation{Department of Physics, California Institute of Technology, Pasadena, California 91125, USA}
	\affiliation{Institute for Quantum Information and Matter, California Institute of Technology, Pasadena CA 91125, USA}

	\begin{abstract}
		Recent experiments on Bernal bilayer graphene (BLG) deposited on monolayer WSe$_2$ revealed robust, ultra-clean superconductivity coexisting with sizable induced spin-orbit coupling. Here we propose BLG/WSe$_2$ as a platform to engineer \emph{gate-defined} planar topological Josephson junctions, where the normal and superconducting regions descend from a common material. 
		More precisely, we show that if superconductivity in BLG/WSe$_2$ is gapped and emerges from a parent state with intervalley coherence, then MZMs can form in the barrier region upon applying weak in-plane magnetic fields. 
		Our results spotlight a potential pathway for `internally engineered' topological superconductivity that minimizes detrimental disorder and orbital-magnetic-field effects. 
	\end{abstract}
	
	\date{\today}
	
	\maketitle
	
	Experimental searches for non-Abelian anyons have to date largely followed two complementary paths.  The first seeks intrinsic realizations of 
	strongly correlated topological phases of matter, most notably non-Abelian fractional quantum Hall states \cite{Nayak2008} and quantum spin liquids \cite{kitaev2006anyons}.  The second endeavors to engineer 
	topological superconductors by interfacing well-understood building blocks---e.g., conventional superconductors and semiconductors---that 
	originate from disparate materials \cite{Fuliang2008, Jaysau2010, Roman2010, Oreg2010, Alicea2010,  Mourik2012, Choy2011, Nadj2013, Pientka2013, DanielLoss2013, stevan2014, Xiaoliang, alicea2012new, beenakker2013search,  Lutchyn2018, Flensberg2021}.  One can, however, contemplate a middle ground between these strategies, wherein phases of matter intrinsic to a \emph{single medium} are leveraged to `internally engineer' topological superconductivity.  Graphene multilayers comprise an attractive platform for the latter approach given their extraordinarily rich and tunable phase diagrams. 
	As proof of concept, Ref.~\onlinecite{Thomson2022} proposed that gate-defined wires judiciously immersed between gapped phases of twisted bilayer graphene~\cite{Cao2018sc, Yankowitz2019, Lu2019, Arora2020, Oh2021}, or it multilayer generalizations~\cite{Park2021, Philip2021, Kim2022, Park2022, Yiran2022}, could realize topological superconductivity without invoking `external' proximity effects (see also Refs.~\onlinecite{Rodan-Legrain2021, deVries2021,Portoles2022, Jaime2021} for related architectures). 
	
	\emph{Untwisted} (i.e., crystalline) graphene multilayers 
	exhibit phase diagrams whose richness and tunability rival that of their twisted counterparts.  Here, applying a perpendicular displacement field $D$ opens a gap at charge neutrality 
	and locally flattens the bands near the Brillouin zone corners---providing a knob to continuously tune the strength of electronic correlations.  Experiments have 
	reported a series of correlation-driven symmetry-broken 
	metallic states together with superconductivity in both Bernal bilayer graphene (BLG) \cite{Zhou2021_BLG,delaBarrera2022,Seiler2022, zhang2023,Holleis2023, LiJia2023} and rhombohedral trilayer graphene~\cite{Zhou2021_cor, Zhou2021_sc}, inspiring various theory proposals for the underlying pairing mechanism~\cite{Berg2021, Chatterjee2022, Cea2022, Szab2022, You2022, Chou2022, YangZhi2022, Patri2022, Curtis2022, Wagner2023, Jimeno-Pozo2022, Qin2022, Lu2022, Dai2022, Dong2022, Shavit2023,Dong2023}.
	In BLG, superconductivity was first observed over a narrow density window in the presence of in-plane magnetic fields $B_\parallel \gtrsim 150$ mT, with a low critical temperature $T_c\approx 30$ mK \cite{Zhou2021_BLG}.  More recent experiments~\cite{zhang2023, Holleis2023} found that placing BLG adjacent to monolayer WSe$_2$ both generates appreciable spin-orbit coupling (SOC) \emph{and} promotes Cooper pairing: Superconductivity appears over a
	broader density window within a symmetry-broken parent metallic phase, with $T_c$ up to hundreds of mK, and without any applied magnetic field. 
	Similar trends have now been observed in several graphene-based systems \cite{Arora2020,  Ruiheng2022,zhang2023}, suggesting a deep connection between SOC and enhanced pairing ~\cite{Curtis2022, Wagner2023, Jimeno-Pozo2022, Ming2023}.
	
	\begin{figure}
		\centering
		\includegraphics[width=\columnwidth]{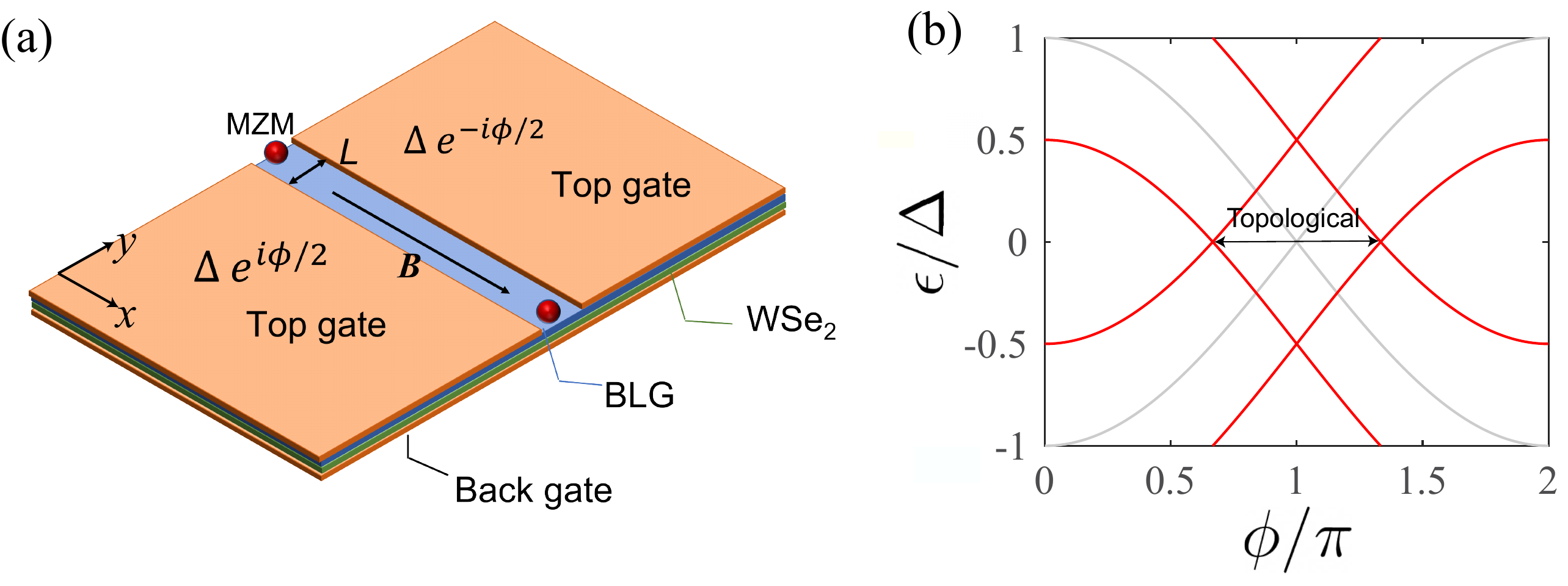}
		\caption{(a) Gate-defined Josephson junction in BLG/WSe$_2$ predicted to host Majorana zero-energy modes (MZMs) near a phase difference $\phi = \pi$ with small in-plane magnetic fields $\bm{B}$.
			(b) Andreev bound states spectrum from Eq.~\eqref{ABS_energy} with Zeeman energy $\tilde{h}=0$ (gray) and $\tilde{h} = 0.5\Delta$ (red), the latter opening a topological regime.}
		\label{fig:fig1}
	\end{figure}
	
	Here we propose BLG/WSe$_2$ as a new platform for internally engineered topological superconductivity. Our proposal is inspired by seminal theory works \cite{Pientka2016, Flensberg2017} which showed that spin-orbit-coupled planar Josephson junctions at a phase difference of $\pi$ can in principle host topological superconductivity at arbitrarily weak magnetic fields.  Numerous experiments have since pursued this approach in junctions fashioned from proximitized heterostructures \cite{Ren2019, Fornieri2019, Ke2019, Dartiailh2021, Banerjee2022}. 
	We show that Josephson junctions \emph{generated solely by electrostatic gating} in BLG/WSe$_2$ [Fig.~\ref{fig:fig1}(a)] can similarly host a topological regime at weak magnetic fields, provided two requirements are satisfied:  Superconductivity native to BLG/WSe$_2$ must exhibit a bulk gap (to ensure well-defined Andreev bound states (ABSs) in the junction) and descend from a symmetry-broken normal state with intervalley coherence (to lift valley degeneracy while maintaining the resonance condition for intervalley Cooper pairing).  At present the superconducting order parameter and normal-state symmetry in BLG/WSe$_2$ remain unknown.  Hartree-Fock treatments do, however, predict that various types of IVC states are energetically competitive in BLG \cite{zhang2023, Ming2023}, rhombohedral trilayer graphene \cite{Chatterjee2022, Wang2023, Zhumagulov2023, koh2023correlated}, and twisted bilayer graphene \cite{Zaletel2020, Yizhang2020, Lianbiao2021, Wagner2022} (see also a recent tensor-network study~\cite{Wang2022}). Moreover, IVC order was recently imaged using STM in monolayer graphene in its zeroth Landau level~\cite{Liu2022, Coissard2022} as well as in twisted graphene bilayers~\cite{Nukolls2023} and trilayers \cite{Hyunjin_2023}.  Turning the problem on its head, one can view our proposal as a transport probe of IVC order in BLG/WSe$_2$.
	
	Aside from eliminating the need for heterostructure engineering, our proposal entails two other key advantages: (1) Superconductivity in BLG/WSe$_2$ occurs deep in the clean limit \cite{Zhou2021_BLG,zhang2023}, thereby mitigating adversarial disorder effects that remain a major obstruction in the pursuit of topological superconductivity and (2) the atomically thin nature of the setup reduces detrimental orbital effects of in-plane magnetic fields~\cite{Anton2016, Banerjee2022, MSgap2022}. 
	
	{\bf \emph{Symmetry-based low-energy description}.}~First we derive a minimal effective Hamiltonian for the valence band of BLG/WSe$_2$ at large displacement fields $D$---where superconductivity emerges.  
	%(See
	%Ref.~\onlinecite{Zhou2021_BLG} and 
	%the Supplementary Material (SM) \cite{Supp} for band-structure details.) 
	As a baseline, Fig.~\ref{fig:fig2}(a) sketches the large-$D$ %free-particle 
	valence band %energies 
	in the absence of SOC.  
	The
	low-energy degrees of freedom carry spin and valley quantum numbers associated with Pauli matrices $s_{x,y,z}$ and $\tau_{x,y,z}$, respectively. The system preserves time-reversal symmetry $\mathcal{T}=i\tau_xs_y \cal{K}$ ($\cal{K}$ denotes complex conjugation) and three-fold rotations $C_{3}=e^{-i\frac{\pi}{3}s_z}$. 
	We also impose the approximate $x\rightarrow -x$ mirror symmetry $M_x=i\tau_xs_x$, even though it is weakly broken by the WSe$_2$ substrate, and (for now) enforce valley conservation.
	
	We express the single-particle Hamiltonian respecting these symmetries as
	\begin{equation}
		H_0(\bm{k})=h_0(\bm{k})+h_{\rm so}(\bm{k}),
		\label{eq_H0}
	\end{equation}
	where $\bm{k}$ denotes momentum measured with respect to the $K$ and $K'$ points. 
	The first term captures the SOC-free band dispersion and can be decomposed as $h_0(\bm{k})=\xi_0(\bm{k})+\xi_1(\bm{k})\tau_z$.
	We take $\xi_0(\bm{k})\approx -
	\mu + t_a\bm{k}^2+t_c\bm{k}^4$ with $\mu$ the chemical potential; the valley-dependent contribution encodes trigonal warping and, to leading
	order in momentum, takes the form $\xi_1(\bm{k})\approx t_b(k_x^3-3k_xk_y^2)$.
	By fitting to the full dispersion~\cite{Supp}, we estimate $t_a= 4$ eV$\cdot$ $a^2$, $t_b=-60$ eV$\cdot$ $a^3$, $t_c=-1500$ eV$\cdot$ $a^4$, with $a=0.246$~nm the lattice constant.
	The second term in Eq.~\eqref{eq_H0} captures SOC in BLG 
	inherited via virtual tunneling to WSe$_2$:
	\begin{equation}
		h_{\rm so}(\bm{k})=\frac{\beta_I}{2}\tau_zs_z+\alpha_R (k_xs_y-k_ys_x).
		\label{eq:soc}
	\end{equation}
	Here, $\beta_I$ and $\alpha_R$ respectively denote Ising and Rashba SOC couplings, whose magnitudes 
	depend on the BLG/WSe$_2$ interface quality 
	and twist angle~\cite{Li2019, David2019, Naimer2021, YangZhi2022}. For example, $\beta_I \sim 0.7-2$ meV was extracted using quantum Hall measurements in different devices ~\cite{Island2019, zhang2023, Holleis2023}. 
	The Rashba scale is harder to directly measure, but can be conservatively estimated~\cite{Supp} as $\alpha_R \sim 1-3\text{meV}\cdot a$.
	
	\begin{figure}
		\centering
		\includegraphics[width=1\linewidth]{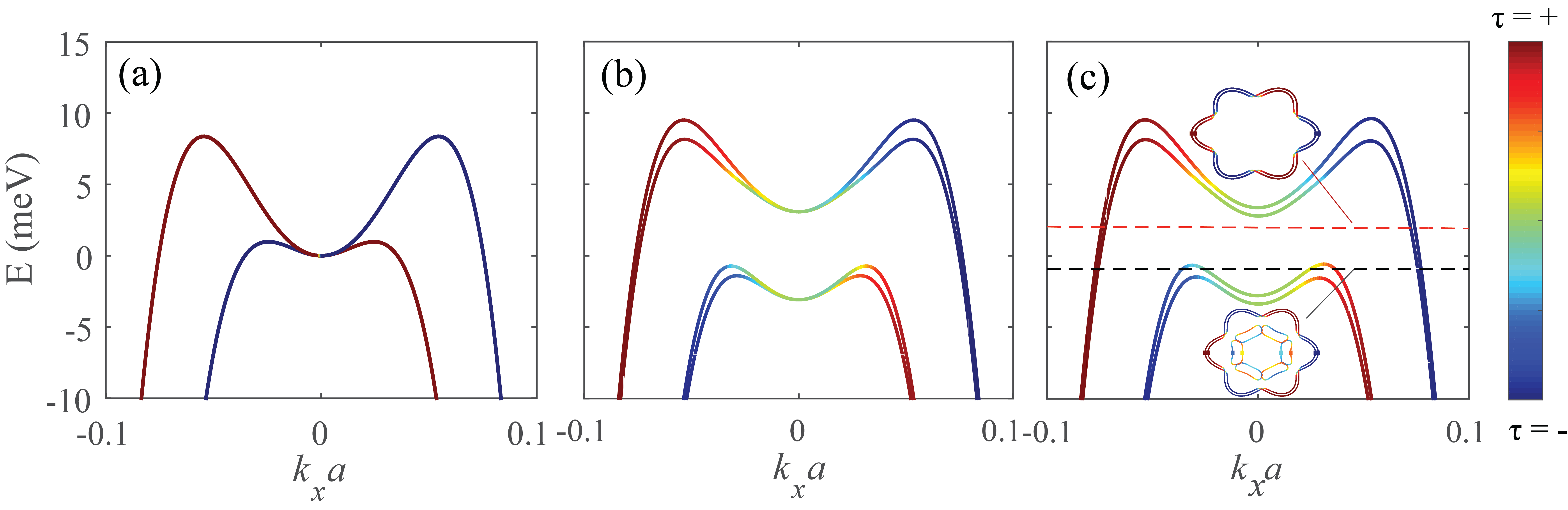}
		\caption{Valence bands of BLG/WSe$_2$ at large displacement fields, calculated from Eq.~\eqref{normal} with $k_y=0$ and $\lambda_1 = 0$.  Parameters are (a) $\beta_I = \alpha_R = \lambda_0 = 0$ and (b,c) $\beta_I=1.4$ meV, $\alpha_R=2$ meV$\cdot a$, $\lambda_0=3$ meV; panel (c) further includes a Zeeman field $h=0.3$ meV (along the $x$-direction). 
			Insets in (c) show the Fermi pockets arising with chemical potentials shown by the red and black dashed lines.  Energy bands are colored according to their valley projection.} 
		\label{fig:fig2}
	\end{figure}
	
	{\bf \emph{Intervalley coherence.}}~Topological superconductivity emerges when an odd number of Fermi surfaces acquire a pairing gap.  To this end, SOC in coordination with a Zeeman field facilitates the removal of spin degeneracy, though valley degeneracy in BLG provides an added obstruction. Coulomb interactions can lift the latter degeneracy by promoting symmetry-breaking within the spin-valley subspace, akin to Stoner ferromagnetism~\cite{Zhou2021_cor, Zhou2021_BLG, zhang2023}. Valley polarized states---in which electrons preferentially populate either the $K$ or $K'$ valley---are antagonistic to Cooper pairing and thus likely irrelevant for the parent state of superconductivity.
	We instead focus on IVC orders, wherein valley degeneracy is lifted via spontaneous coherent tunneling between $K$ and $K'$; 
	such states more naturally host superconductivity since the resonance condition for intervalley Cooper pairing can persist. Table I in the SM \cite{Supp} classifies possible IVC orders.
	While we expect that our proposal holds for generic IVC states, so long as they support gapped zero-momentum Cooper pairing, we focus 
	for concreteness
	on the subset that preserves $\mathcal{T}$, $C_3$, and $M_x$.  To linear order in momentum, the corresponding IVC order parameter can be expressed as
	\begin{equation}
		\Delta_{\rm IVC}(\bm{k})= \lambda_0\tau_x+\lambda_{1}\tau_x (k_xs_y-k_ys_x),
		\label{eq_IVC_triplet}
	\end{equation}
	where $\lambda_0$ describes a spin- and momentum-independent contribution and $\lambda_1$ encodes a spin-valley-orbit coupling. 
	
	Supplementing Eq.~\eqref{eq_H0} with both the above IVC order parameter and an in-plane Zeeman field $\bm{h}$ yields the putative normal-state Hamiltonian
	\begin{equation}
		H(\bm{k})= H_0(\bm{k})+\Delta_{\rm IVC}(\bm{k}) + \bm{h\cdot s}  \label{normal}
	\end{equation}
	that can exhibit fully lifted spin and valley degeneracies.
	Figure~\ref{fig:fig2}(b,c) sketches the band structure evolution upon (b) turning on SOC and IVC order and then (c) further adding a Zeeman field. 
	
	When the chemical potential intersects only the upper pair of bands [red dashed line in Fig.~\ref{fig:fig2}(c)],
	one can further distill the model by projecting out the lower, inert bands---yielding an effective Hamiltonian
	\begin{equation}
		\tilde H(\bm{k})=\xi_0(\bm{k})+\tilde\beta_I (k_x^3-3k_xk_y^2)\sigma_z+\tilde \alpha_R (k_x\sigma_y-k_y\sigma_x)+\bm{\tilde{h}\cdot\sigma}
		\label{eff_Ham}
	\end{equation}
	valid in the regime $h \ll \sqrt{\lambda_0^2 + \frac{\beta_I^2}{4}}$.
	Here $\sigma_{x,y,z}$ are Pauli matrices acting in the low-energy subspace and
	\begin{equation}
		\tilde \beta_I=\frac{t_b\beta_I}{2\sqrt{\lambda_0^2+\frac{\beta_I^2}{4}}}, ~\tilde\alpha_R=\frac{\alpha_R\lambda_0}{\sqrt{\lambda_0^2+\frac{\beta_I^2}{4}}}+\lambda_1,~\bm{\tilde{h}}=\frac{\bm{h}\lambda_0}{\sqrt{\lambda_0^2+\frac{\beta_I^2}{4}}}.
		\label{parameters_projected}
	\end{equation}
	Equation~\eqref{eff_Ham} represents a two-band model for fermions with cubic-in-momentum Ising SOC and linear-in-momentum Rashba SOC. Note that IVC order suppresses $\tilde \beta_I$, but enhances $\tilde \alpha_R$ and $\tilde h$   
	through a linear coupling to the bare Zeeman field $h$ and Rashba SOC $\alpha_R$, which would only be quadratic when $\lambda_0 = 0$. The spin-valley-orbit term
	$\lambda_1$ simply contributes to the effective Rashba coupling; we thus set $\lambda_1 = 0$ hereafter.
	Parameter renormalizations in Eq.~\eqref{parameters_projected} may be relevant for the unconventional Pauli-limit-violation trends observed in Ref.~\onlinecite{zhang2023}. 
	
	{\bf \emph{Topological Josephson junctions.}}~We now incorporate superconductivity, assuming for simplicity an s-wave order parameter---though we stress that our scheme readily extends to more exotic pairings provided they generate a bulk gap. 
	The corresponding Bogoliubov-de Gennes (BdG) Hamiltonian
	reads
	\begin{equation}
		\tilde H_{\rm BdG}(\bm{k})=\begin{pmatrix}
			\tilde H(\bm{k})& i\Delta \sigma_y\\
			-i\Delta\sigma_y&-\tilde H^*(-\bm{k})
		\end{pmatrix}
		\label{Mean_field}
	\end{equation} 
	with $\Delta$ the pairing amplitude.
	To set the stage, we first consider a simple case where $\xi_0(\bm{k})\approx -\bm{k}^2/2m-\mu$ and $\tilde \beta_I k_F^3 \ll \tilde \alpha_R k_F$ ($m$ is an effective mass and $k_F$ is the Fermi momentum). %Projecting the pairing term to the low-energy basis of Eq.~\ref{eff_Ham}, we have $\Delta=i\sigma_y$. 
	Equation~\eqref{Mean_field} then maps to the Hamiltonian for a proximitized Rashba-coupled 2D electron gas under in-plane magnetic fields---exactly the ingredients required to create topological superconductivity in planar Josephson junctions~\cite{Pientka2016, Flensberg2017}.  
	
	Next we consider a Josephson junction, with phase difference $\phi$, formed by  gate-tuning a barrier region of length $L$ into a 
	normal phase with $\Delta=0$. The magnetic field is oriented parallel to the junction (along the $x$ direction), which is optimal for accessing the topological regime~\cite{Pientka2016, Flensberg2017}. Within the barrier the relevant Fermi velocity $v_F$ arises from the large pockets in Fig.~\ref{fig:fig2}(c); band-structure estimates give $v_F \sim 5\times 10^5$ m/s. For reasonable junction lengths $L \sim 50 - 200$~nm, the corresponding Thouless energy $E_T=\frac {\pi v_F}{2L} \sim 0.8 - 3$~meV greatly exceeds both the pairing energy $\Delta$ and 
	renormalized Zeeman energy $\tilde h$.
	We therefore assume the short-junction limit $E_T\gg \Delta, \tilde{h}$ below.
	
	Topological phase transitions are determined by computing the ABS spectrum at $k_x = 0$ using the standard scattering matrix method (for details see SM).  In the absence of normal reflections at the interfaces, the $k_x = 0$ ABS energies take the simple form
	\begin{equation}
		\epsilon_{\pm,\eta}(\phi)=\eta\tilde{h}\pm \Delta\cos(\phi/2),
		\label{ABS_energy}
	\end{equation}
	where $\eta = \pm$ is a pseudospin label.
	Figure~\ref{fig:fig1}(b) plots these energies at $\tilde{h} = 0$ and  $\tilde{h}=0.5\Delta$. In the zero-field limit, both $\eta = \pm $ branches become gapless at a phase difference $\phi = \pi$. Turning on the in-plane Zeeman field
	shifts the gap closing points to $\phi \neq \pi$, thereby opening up a topological regime. The above features are consistent with the results of Refs.~\cite{Pientka2016, Flensberg2017}, except that we consider a uniform Zeeman energy in the barrier and superconducting regions. 
	The $\tilde h$ contribution in $\epsilon_{\pm,\eta}$ descends from the Zeeman energy in the superconductors---which produces non-degenerate $k_x = 0$ energies at $\phi = 0$. 
	
	{\bf \emph{Numerical phase diagram.}}~We now verify the above physical picture via a numerical calculation of the ABSs in a more realistic BLG/WSe$_2$ model. We rewrite the four-band Hamiltonian capturing the low-energy bands, Eqs.~(\ref{normal}) and~(\ref{Mean_field}), as a tight-binding model in the $y$ direction for a given $k_x$
	(see SM Sec.~V). To emulate experiments, where quantum oscillations reveal that superconductivity occurs in a regime with both large and small pockets~\cite{zhang2023}, we fix the chemical potential in the superconducting regions to $\mu_2=-1$ meV
	[black dashed line in Fig.~\ref{fig:fig2}(c)].
	The chemical potential $\mu_1$ in the barrier can be tuned independently, including to a regime with only large pockets (where the effective two-band model applies).
	For concreteness, unless specified otherwise we set $\mu_1 = 2$ meV, $\beta_I=1.4$ meV, $\alpha_R = 2$ $\text{meV} \cdot a$, $\lambda_0=3$ meV, $\lambda_1 = 0$, and $\Delta=0.1$ meV;
	under BCS weak-coupling assumptions the latter corresponds to $T_c \sim 600$ mK.  Note that this choice of $\lambda_0$ and $\beta_I$ gives $\tilde h \approx h$.

	The ABS spectrum follows by diagonalizing the 1D tight-binding model for each $k_x$. Figure~\ref{fig:fig3}(a,b) presents the $k_x = 0$ energies versus $\phi$ for (a) $h=0$ and (b) $h= 0.5\Delta$.  In (a), lifting of valley degeneracy by IVC order leaves two nearly degenerate ABSs with a small splitting (away from $\phi = 0,\pi$).
	Panel (b) shows that a Zeeman field nucleates a topological region centered around $\phi = \pi$, separated from the trivial regions by a $k_x = 0$ gap closure.
	These observations agree well with simpler two-band-model results from Fig.~\ref{fig:fig1}(b); indeed the bound states continue to be well-captured by Eq.~\eqref{ABS_energy} [see dashed lines in Fig.~\ref{fig:fig3}(b)].  Figure \ref{fig:fig3}(c) shows the broader phase diagram, obtained from the $k_x = 0$ gap ($E_g$) illustrated by the colormap, versus $h/\Delta$ and $\phi$. In the topological phase, the minimal gap $E_{g,m}$ typically appears at finite $k_x$ and is maximized when SOC effects are dominated by the effective Rashba contribution (see SM Sec.~VI). The topological phase transition lines closely emulate the curves $\tilde{h}=\Delta|\cos(\phi/2)|$ predicted by Eq.~\eqref{ABS_energy}. 
	For the $\pi$-junction, topological superconductivity sets in at arbitrarily weak Zeeman fields---though more generally normal reflections at the interfaces push the required Zeeman field to a finite value (see SM Sec.~III) \cite{Pientka2016}.
	As the in-plane field increases, the topological phase persists until the superconducting regions become gapless, roughly when
	the renormalized Zeeman energy $\tilde{h} \sim \Delta$.
	
	\begin{figure}
		\centering
		\includegraphics[width=1\linewidth]{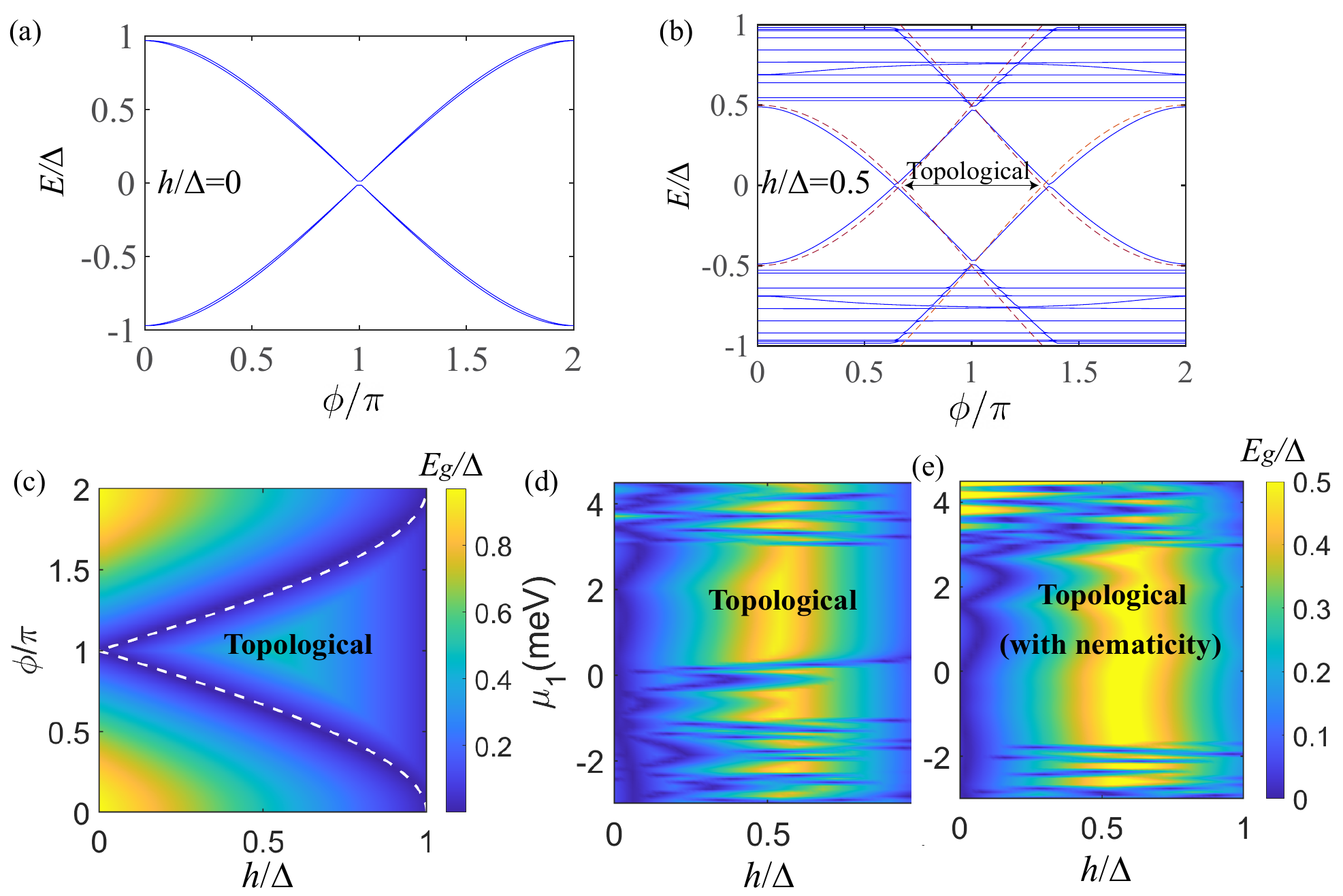}
		\caption{
			(a,b) ABS energies $E$ at $k_x = 0$ 
			as a function of $\phi$, obtained from a realistic tight-binding model with Zeeman energy (a) $h = 0$ and (b) $h =  0.5\Delta$. Dashed lines in (b) trace the analytical result Eq.~(\ref{ABS_energy}) obtained in the short junction limit. (c-e) Phase diagrams as a function of (c) $h, \phi$ and (d,e) $h,\mu_1$.  In (c) the dashed white line indicates $h=\Delta|\cos(\phi/2)|$, which roughly captures the topological phase boundary.  Panel (e) is the same as (d), but with nematicity phenomenologically incorporated. Data correspond to $L=158$ nm and $\phi=\pi$. %$\mu_1 = 2$ meV, $\mu_2 = -1$ meV, $\beta_I=0.7$ meV, $\alpha_R=2$ $\text{meV} \cdot a$,
			%$\mu_2=-1$ meV (black dashed line in (b)), $L\approx 150$nm, 
			%     $\lambda_0=3$ meV, and $\phi=\pi$.  
		}
		\label{fig:fig3}
	\end{figure}
	
	Figure~\ref{fig:fig3}(d) reveals the dependence of $E_g$ on $h/\Delta$ and the barrier chemical potential $\mu_1$ at $\phi = \pi$.  A robust topological region occurs for $\mu_1 \sim 0-3$ meV, where the barrier hosts only two large hole pockets. Outside of this range, additional 
	small pockets arise---see Fig.~\ref{fig:fig2}(c)---that engender frequent topological phase transitions.  Quantum oscillations, however, indicate that the number of small pockets above $T_c$ is smaller than the six naively predicted by band theory~\cite{zhang2023}, which was suggested to arise from electronic nematicity~\cite{Jung2015, Dong2021}. Including nematicity in our calculation (see SM Sec.~IV), we find that the overall features of the phase diagram persist, while
	%for certain orientations of the nematic order parameter 
	%[as in Fig.~\ref{fig:fig3}(e)] 
	the extent of the robust topological region
	increases due to fewer ``polluting" low-energy states. 
	
	\begin{figure}
		\centering
		\includegraphics[width=1\linewidth]{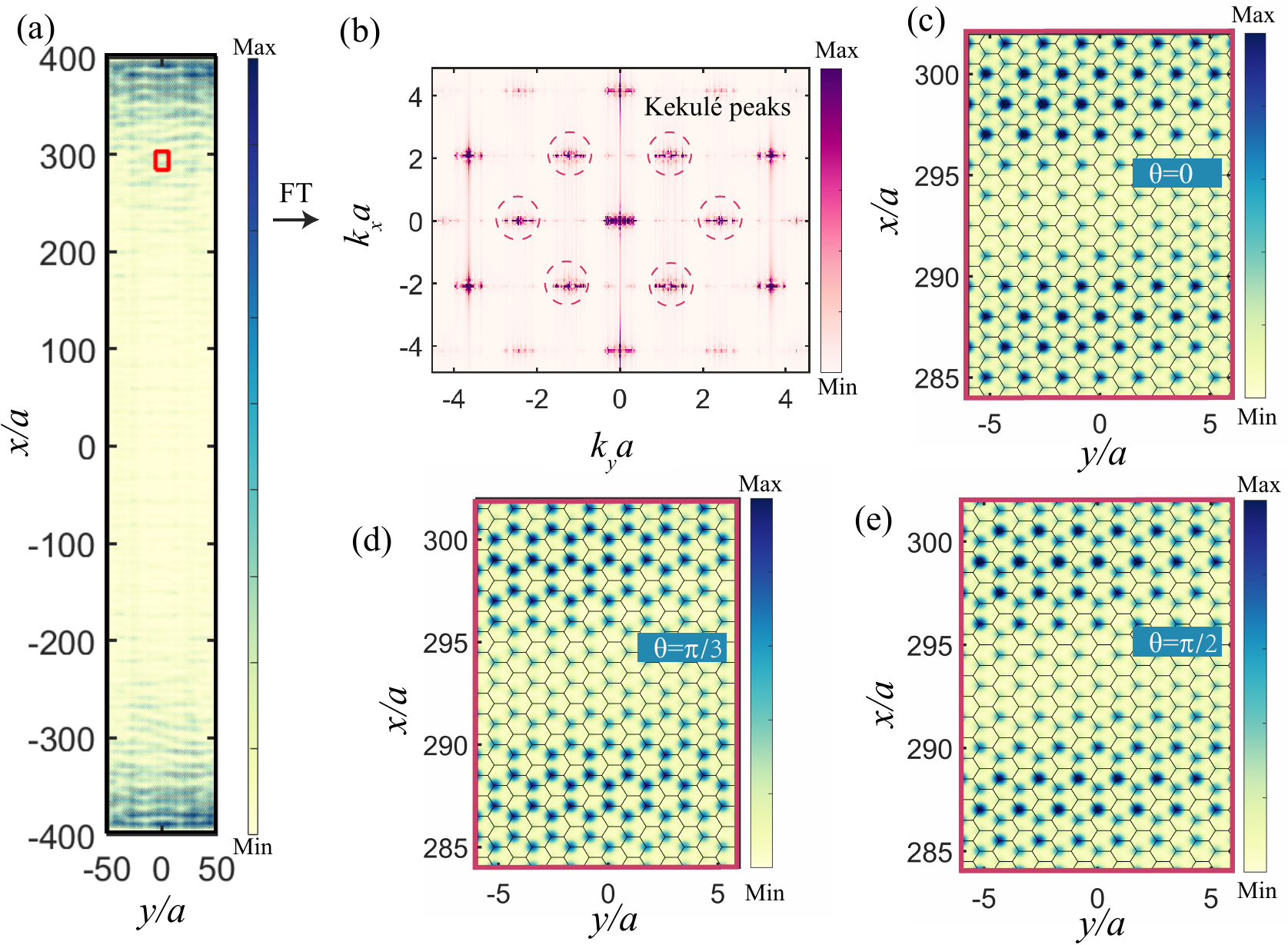}
		\caption{(a) Atomically resolved density of states 
			for the MZM wavefunction in the junction, assuming Kekulé angle $\theta=0$.
			(b) Fourier transform (FT) of the data in (a) revealing Kekulé peaks resulting from IVC-induced atomic-scale reconstruction. 
			(c-e) Zoom-in of the red rectangular region in (a) showing the evolution of the symmetry-breaking pattern with different Kekulé angles.
			Data correspond to $L=160a$, $h/\Delta=0.8$ and $\phi=\pi$.}
		\label{fig:fig5}
	\end{figure}
	
	{\bf \emph{Atomically resolved MZM wavefunctions.}}~Topological superconductivity acquires a novel fingerprint in our setup: The essential ingredient of IVC order generates atomic-scale translation symmetry breaking with an enlarged $\sqrt{3}\times \sqrt{3}$ Kekulé supercell, 
	which manifests directly in the MZM wavefunctions. 
	Figure~\ref{fig:fig5}(a) illustrates the 
	atomically resolved MZM 
	density of states in the barrier (see SM~\cite{Supp} Sec.~VIII);  the usual exponential localization and Friedel-like oscillations are evident on the scale shown. Fourier transforming the data
	reveals 
	characteristic 
	momentum-space peaks [Fig.~\ref{fig:fig5}(b)] associated with the %IVC-induced 
	Kekulé supercell, while
	Fig.~\ref{fig:fig5}(c) zooms in on the red rectangular
	region from (a) and clearly shows 
	the corresponding atomic-scale ordering. 
	Contrary to the familiar `bond-centered' Kekulé patterns observed in monolayer graphene~\cite{Liu2022, Coissard2022},
	here symmetry breaking manifests primarily on sites due to sublattice and layer polarization generated at large $D$ fields.
	Figure~\ref{fig:fig5}(d,e) explores different Kekulé angles $\theta$, obtained by replacing the $\tau_x$ %IVC 
	order parameter considered thus far with $\cos(\theta) \tau_x + \sin(\theta) \tau_y$.
	The resulting Kekulé pattern intimately relates to $\theta$, allowing experimental characterization of IVC order via the MZMs.  
	
	STM measurements that resolve zero-bias peaks without the atomic-scale structure predicted here could arise from %low-energy 
	trivial ABSs originating from disorder \cite{Liujie2012,Bargrets2012, Das2021} or inhomogeneities near the barrier ends \cite{Tewari2018, Frolov2019, Tewari2019} (see Refs.~\cite{Prada2020,Flensberg2021} for reviews).
	Conversely, and more definitively,
	observing localized zero modes at both junction ends that appear near $\phi = \pi$ 
	\emph{and} exhibit
	Kekulé order would strongly support topological superconductivity arising from an IVC normal state.
	
	{\bf \emph{Discussion and Outlook.}}~We proposed a route to one-dimensional topological superconductivity where all required ingredients---SOC, Cooper pairing, and the ability to fabricate planar Josephson junctions---appear natively in a \emph{single} BLG/WSe$_2$ platform. Our proposal relies on two reasonable but so far untested assumptions: a gapped superconducting phase and a normal parent state with intervalley coherence. The ultra-clean nature of superconductivity in BLG~\cite{Zhou2021_BLG, zhang2023}, with an electronic mean-free path far exceeding the coherence length, presents an enormous virtue that 
	potentially circumvents 
	disorder effects that plague proximitized nanowires \cite{Liujie2012,Bargrets2012, Das2021} and Josephson junctions~\cite{Fornieri2019, Ren2019, Banerjee2022}. Weak disorder arising, e.g., from imperfections in the geometry of gates defining the junction, can even enhance the robustness of the topological phase by decreasing the Majorana localization length~\cite{Ady2019}. %Such enhancement arises when large-momentum electrons propagating almost parallel to the junction are scattered by disorder, effectively enhancing the (internal) superconducting proximity effect within the junction
	A more controllable route to the same goal consists of gate-defining the junction in a zigzag geometry designed to enhance Andreev reflections~\cite{Laeven2020}.
	
	Our proposal readily generalizes to more exotic order parameters (see SM Table I). If either the normal or superconducting state spontaneously breaks time-reversal symmetry, topological superconductivity could arise without an applied magnetic field~\cite{Thomson2022}. For example, a spin-nematic IVC state described by a $\tau_x s_x$ order parameter generates an effective Zeeman field when projected to the low-energy subspace of Eq.~\eqref{eff_Ham}. 
	We also expect our proposal to be relevant for a broader family of graphene-based structures. Substantial efforts have been devoted to gate-defined wires and Josephson junctions in 
	twisted bilayer graphene \cite{Rodan-Legrain2021, deVries2021, Portoles2022, Jaime2021, Yingming2022,Jinxin2022, Sainz2022, Thomson2022}, which also enjoys exquisite gate tunability but suffers from more prevalent disorder \cite{Uri2020, Wilson2020}. Rhombohedral trilayer graphene offers a cleaner platform for gate-tunable correlated states and superconductivity~\cite{Zhou2021_sc, Zhou2021_cor}---thus presenting another interesting medium for future exploration along these lines.   
	
	{\bf \emph{Acknowledgements.}}~ We are grateful to Andrey Antipov, Cory Dean, Cyprian Lewandowski, Alex Thomson, and Yiran Zhang for enlightening discussions. Y.M.X.~acknowledges the support of HKRGC through PDFS2223-6S01. \'E.L.-H.~was supported by the Gordon and Betty Moore
	Foundation’s EPiQS Initiative, Grant GBMF8682. J.A. was supported by the Army Research Office under Grant Award W911NF-17-1-0323; the Caltech Institute for Quantum Information and Matter, an NSF Physics Frontiers Center with support of the Gordon and Betty Moore Foundation through Grant GBMF1250; and the Walter Burke Institute for Theoretical Physics at Caltech.  The U.S. Department of
	Energy, Office of Science, National Quantum Information Science Research Centers, Quantum Science Center supported the symmetry-based analysis of this work.  S.N.P.~acknowledges support of Office of Naval Research (grant no. N142112635) and NSF-CAREER (DMR-1753306) programs. Work at UCSB was supported by the U.S. Department of Energy (Award No. DE-SC0020305).

	\bibliography{Reference}
	
	\clearpage

	\onecolumngrid
	\begin{center}
		\textbf{\large Supplementary Material for  \lq\lq{}Gate-defined topological Josephson junctions in Bernal bilayer graphene \rq\rq{}}\\[.2cm]
		Ying-Ming Xie,$^{1,2,3}$ \'Etienne Lantagne-Hurtubise,$^{2,3}$, Andrea F.~Young,$^{4}$ Stevan Nadj-Perge,$^{4,3}$, Jason Alicea,$^{2,3}$ \\[.1cm]
		{\itshape ${}^1$Department of Physics, Hong Kong University of Science and Technology, Clear Water Bay, Hong Kong, China}\\
		{\itshape ${}^2$Department of Physics, California Institute of Technology, Pasadena, California 91125, USA}\\
		{\itshape ${}^3$Institute for Quantum Information and Matter, California Institute of Technology, Pasadena CA 91125, USA}\\
		{\itshape ${}^4$ Department of Physics, University of California at Santa Barbara, Santa Barbara CA 93106, USA} \\
		{\itshape ${}^5$T. J. Watson Laboratory of Applied Physics, California Institute of
			Technology, 1200 East California Boulevard, Pasadena, California 91125, USA}\\[1cm]
	\end{center}
	\setcounter{equation}{0}
	\setcounter{section}{0}
	\setcounter{figure}{0}
	\setcounter{table}{0}
	\setcounter{page}{1}
	\renewcommand{\theequation}{S\arabic{equation}}
	\renewcommand{\thetable}{S\arabic{table}}
	\renewcommand{\thesection}{\Roman{section}}
	\renewcommand{\thefigure}{S\arabic{figure}}
	\renewcommand{\bibnumfmt}[1]{[S#1]}
	\renewcommand{\citenumfont}[1]{#1}
	\makeatletter
	
	\twocolumngrid
	
	\maketitle

	\section{Minimal model for BLG at finite displacement fields}
	
	The four-band continuum model  describing BLG under a perpendicular displacement field $D$ can be written as 
	\begin{equation}
		h_\tau(\bm{k})=\begin{pmatrix}
			u/2&v_0\Pi^{\dagger}&-v_4\Pi^{\dagger}&-v_3\Pi\\
			v_0\Pi&\Delta'+u/2&\gamma_1&-v_4\Pi^{\dagger}\\
			-v_4\Pi&\gamma_1&\Delta'-u/2& v_0\Pi^{\dagger}\\
			-v_3\Pi^{\dagger}&-v_4\Pi&v_0\Pi& -u/2
		\end{pmatrix}.
		\label{app_eq_continuum}
	\end{equation} 
	On the left side, $\tau=\pm 1$  indicates the two valleys $\pm\bm{K}=(\pm  \frac{4\pi}{3a},0)$, with $a$ the lattice constant, and $\bm{k}$ denotes the momentum measured with respect to $\pm \bm{K}$.
	On the right side, $\Pi=(\tau k_x+ik_y)$ specifies the momentum; the basis is $\psi_{\tau}(\bm{k})=(\psi_{\tau, A_1}(\bm{k}), \psi_{\tau, B_1}(\bm{k}), \psi_{\tau, A_2}(\bm{k}),\psi_{\tau, B_2}(\bm{k}) )$  with $A,B$ labeling the sublattice degree of freedom and $1,2$ labeling the layer; and $v_j\equiv \frac{\sqrt{3}}{2}a\gamma_j$, $\gamma_1$, $\Delta'$, and $u$ are band-structure parameters.  Specifically, $\gamma_j$'s encode various hopping processes, $\Delta'$ is an on-site potential difference resulting from the stacking, and $u = -d_\perp D/\epsilon_{\rm BLG}$ is the energy difference between the two layers caused by the perpendicular displacement field $D$ ($d_\perp = 0.33$ nm is the interlayer distance and $\epsilon_{\rm BLG} \approx 4.3$ the relative permittivity of BLG.)
	
\begin{figure}
		\centering
		\includegraphics[width=1\linewidth]{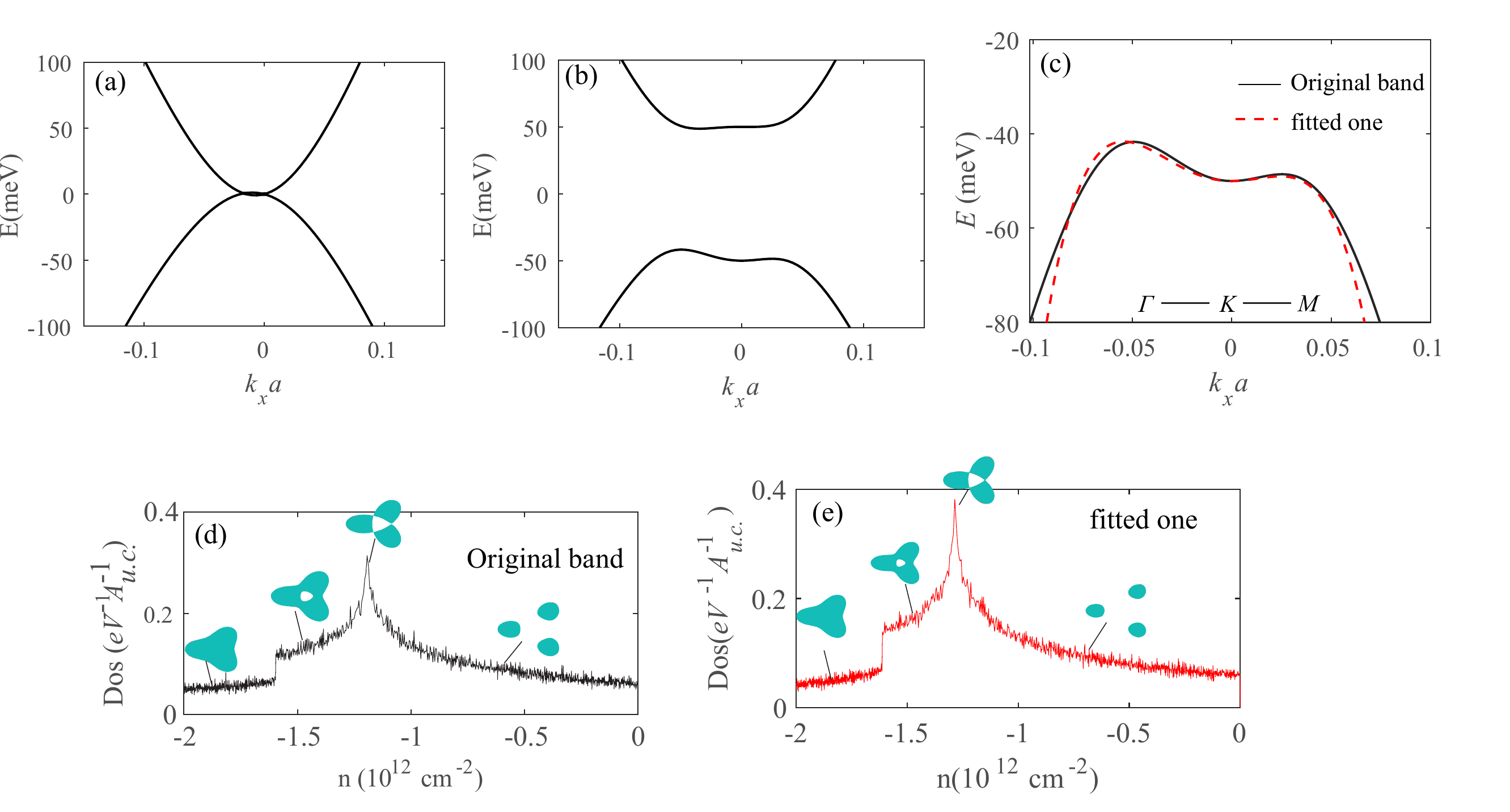}
		\caption{(a,b) Low-energy bands of BLG in valley $+\bm K$ (a) without any displacement field and (b) with a displacement field giving $u=100$ meV. (c) Valence-band  energy $E$ versus $k_x$ in valley $+\bm K$ with $k_y = 0$ and a non-zero displacement field.  Solid black line shows the dispersion obtained from the four-band model $h_+({\bm k})$, while the dashed red line is a fit using the minimal model $h_0({\bm k})$.  (d,e) Density of states (Dos) versus hole density obtained from (d) the four-band model and (e) the minimal model, using  temperature $T=10$ mK for broadening.  The insets show the topography of Fermi contours at various hole densities.}
		\label{fig:figs1}
	\end{figure}

	Parameters recovering first-principles band structures are given in Ref.~\onlinecite{MacDonald2014}: intralayer nearest-neighbor hopping is $\gamma_0= 2.61$ meV; interlayer hoppings are $\gamma_1=361$ meV, $\gamma_3=283$ meV, $\gamma_4= 138$ meV; and the onsite potential difference is $\Delta'=15$ meV.  In recent experiments, the $D$ field yielding optimal superconductivity falls in the range of about  $1\sim 1.3$ V/nm, corresponding to an energy difference $u \approx 80$-$100$ meV~\cite{zhang2023, Island2019}. We fix $u=100$ meV in our simulations with $D \neq 0$.
	Figure~\ref{fig:figs1} illustrates the BLG band structures at (a) zero and (b) nonzero displacement field.   
	
	In the main text, we proposed a minimal model for the valence band that captures the low-energy dispersion for hole-doped BLG near the Fermi energy in the presence of strong displacement fields:
	\begin{equation}
		h_0(\bm{k})=-\mu + t_a\bm{k}^2+t_c\bm{k}^4+t_b (k_x^3-3k_xk_y^2)\tau_z.
	\end{equation}
	To estimate the effective parameters $t_{a,b,c}$, we compare  the original bands obtained from $h_{\tau}(\bm{k})$ with those fitted by the proposed minimal model $h_0(\bm{k})$.  We find that the dispersions roughly match with $t_a= 4$ eV$\cdot$ $a^2$, $t_b=-60$ eV$\cdot$ $a^3$, $t_c=-1500$ eV$\cdot$ $a^4$; see the energies in Fig.~\ref{fig:figs1}(c) as well as the density of states and Fermi contours in Fig.~\ref{fig:figs1}(d,e).
	
	\section{Spin-orbit coupling in the minimal model}

	\begin{figure*}
		\centering
		\includegraphics[width=\linewidth]{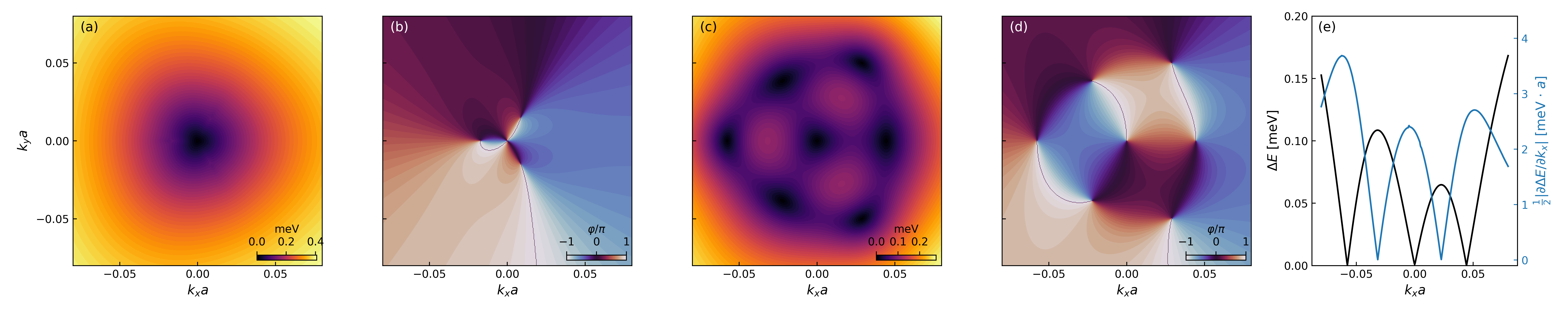}
		\caption{Rashba SOC effects in the top-most valence band of BLG, in the absence of Ising SOC ($\lambda_I=0$). We take $\lambda_R = 1$ meV in Eq.~\eqref{app_eq_SOC}, on the lower end of the experimentally reported range~\cite{Wang2016, Yang2017, Island2019, Wang2019, Amann2022}. Panels (a,b) correspond to BLG/WSe$_2$ at zero displacement field $D$, while panels (c-e) use $D \sim 1.3$ V/nm (corresponding to interlayer potential $u=100$ meV).
			The energy difference $\Delta E$ between the Rasbha spin-split valence bands for momenta near the $\bm K$ point appears in (a,c); the associated spin winding for the top-most spin-split valence band, $\varphi \equiv \text{arg}(\langle s_x \rangle + i \langle s_y \rangle)$, appears in (b,d). (e) Black line: one-dimensional cut through panel (c) for $k_y = 0$. Blue dots: derivative of the spin splitting $\Delta E$ along $k_x$ (for $k_y=0$), which gives a proxy for the parameter $\alpha_R$ defined in the main text [see Eq.~\eqref{eq:soc}].}
		\label{fig_SM_SOC}
	\end{figure*}
	
	At the level of the four-band continuum model, Eq.~\eqref{app_eq_continuum}, the SOC terms appear as~\cite{Gmitra2016, Gmitra2017, Island2019, zhang2023}
	\begin{equation}
		h_{\rm{SOC}, \tau} (\bm{k}) = \mathcal{P}_1 \left[ \frac{\lambda_I}{2} \tau s_z + \frac{\lambda_R}{2} \left( \tau \zeta_x s_y - \zeta_y s_x \right) \right]. 
		\label{app_eq_SOC}
	\end{equation} 
	Here $\zeta$ and $s$ are Pauli matrices acting on the sublattice ($A,B$) and spin degree of freedom, respectively, and $\mathcal{P}_1 = \text{diag}[1, 1, 0, 0]$ projects onto the top layer in the basis used to define Eq.~\eqref{app_eq_continuum}.
	
	Ising and Rashba SOC terms behave very differently within the valence bands of interest.
	At least for ${\bm k}$ near zero, the BLG valence bands become strongly layer and sublattice polarized by a strong $D$ field. This polarization negligibly impacts $\lambda_I$ but suppresses the effects of Rashba SOC due to the latter's off-diagonal structure in sublattice space.  Hence, Rasbha coupling primarily manifests via virtual excitations to remote bands, which 
	can be understood, e.g., using perturbation theory in the limit where trigonal warping is neglected and one considers a theory with a quadratic band touching at $D=0$ (see Ref.~\onlinecite{Zaletel2019}). In reality, trigonal warping splits the band touching into four Dirac points at $D=0$, around which the spins wind due to Rashba SOC; see Fig.~\ref{fig_SM_SOC}(a) and (b). With a strong $D$ field, the Rashba spin-splitting is suppressed and its texture becomes more complicated; see Fig.~\ref{fig_SM_SOC}(c) and (d). The scale of the Rashba spin splitting  strongly depends on whether the Fermi surface (in a given valley) comprises three small pockets or a single large pocket. Figure~\ref{fig_SM_SOC}(e) shows a one-dimensional cut of the Rashba spin splitting along $k_x$ with $k_y = 0$. As a proxy for the $\alpha_R$ parameter defined in Eq.~\eqref{eq:soc}, (one half of) the derivative of the Rashba spin-splitting as a function of $k_x$ is shown in Fig.~\ref{fig_SM_SOC}(e); note that the spin splitting will be \emph{twice} the energy scale $\alpha_R k_x$. In the big Fermi pockets where $k_x a \sim 0.06$, we find that a typical scale for $\alpha_R \sim(1-3) \times \lambda_R \cdot a$. Conservatively estimating $\lambda_R \sim 1$ meV (on the lower end of the experimentally reported range~\cite{Wang2016, Yang2017, Island2019, Wang2019, Amann2022}), we then obtain $\alpha_R \sim 1-3 ~\text{meV}\cdot a$, corresponding to the range specified in the main text.
	
	\begin{table*}
		\caption{Classification of possible translation-invariant IVC order parameters, and other types of order parameters for comparison, based on time-reversal $\mathcal{T}$, $C_3$ rotation, mirror operation $M_x=i\tau_x s_x$, and  $U_{v}(1)$ rotation symmetries.}  
		\begin{tabular}{|c|c|c|c|c|c|c|}
			\hline\hline
			$\Delta_n$& $\mathcal{T}=i\tau_xs_yK$& $C_3=e^{-i\frac{\pi}{3}s_z}$ & $M_x=i \tau_x s_x$& $U_v(1)=e^{i\varphi\tau_z}$ & comment \\\hline$\tau_x(s_x,s_y)$&$-1$&$(x,y)$
			& $(+1,-1)$&no &nematic IVC \\\hline
			$\tau_y (s_x,s_y)$&$-1$&$(x,y)$&$(-1,+1)$&no& nematic IVC \\\hline
			$\tau_x s_0$&$+1$&$+1$&$+1$&no& IVC \\\hline
			$\tau_xs_z$&$-1$&$+1$&$-1$&no& spin-valley intertwined IVC \\\hline
			$\tau_ys_0$&$+1$&$+1$&$-1$&no& IVC \\\hline
			$\tau_ys_z$&$-1$&$+1$&$+1$&no& spin-valley intertwined IVC \\\hline
			$\tau_z(s_x,s_y)$&$+1$&$(x,y)$&$(-1,+1)$&yes& nematic valley polarized \\\hline
			$\tau_zs_0$&$-1$&$+1$&$-1$&yes& valley polarized \\\hline
			$\tau_zs_z$&$+1$&$+1$&$+1$&yes& spin-valley polarized \\\hline
			$\tau_0(s_x,s_y)$&$-1$&$(x,y)$&$(+1,-1)$&yes& nematic spin-polarized \\\hline
			$\tau_0s_z$&$-1$&$+1$&$-1$&yes& spin polarized \\\hline
			$(k_x,k_y)\tau_x(s_x,s_y)$&$+1$&$(x,y)\times(x,y)$
			&$(-1,+1)\times (+1,-1)$&no & spin-orbit-valley intertwined IVC  \\\hline
			$(k_x,k_y)\tau_y (s_x,s_y)$&$+1$&$(x,y)\times(x,y)$&$(-1,+1)\times (-1,+1)$&no & spin-orbit-valley intertwined IVC \\\hline
			$(k_x,k_y)\tau_y s_z$&$+1$&$(x,y)$&$(-1,+1)$&no& spin-orbit-valley intertwined IVC \\\hline
			$(k_x,k_y)\tau_xs_z$&$+1$&$(x,y)$&$(+1,-1)$&no& spin-orbit-valley intertwined IVC \\\hline
		\end{tabular}
		\label{ref:table}
	\end{table*}

	\section{Determining the topological region with the  scattering matrix method}
	
	The solvable toy model for the planar Josephson junction described below Eq.~\eqref{Mean_field} of the main text uses a piecewise-constant valence-band Hamiltonian
	\begin{align}
		&\tilde H_{\rm eff}(k_x,y)=\frac{\partial_y^2-k_x^2}{2m}-\mu(y)+\tilde \alpha_R (k_x\sigma_y+i\partial_y \sigma_x)+ \tilde{h}\sigma_x
	\end{align}
	with
	\begin{align}
		\mu(y)=\mu_1\theta (L/2-|y|)+\mu_2\theta (|y|-L/2).\label{chemical}
	\end{align}
	Above, $\theta$ is a step function and $\mu(y)$ captures the chemical potential profile across the device.  For the pairing potential we take 
	\begin{equation}
		\Delta(y)=\Delta  e^{i\text{sgn}(y)\phi/2}\theta(|y|-L/2)
		\label{pairing}.
	\end{equation}
	The phase difference $\phi$ between the two superconductors can be controlled by applying current or magnetic flux through a loop connected to the junction \cite{Fornieri2019, Ren2019}.
	
	The onset of topological superconductivity in the barrier is most easily diagnosed by studying the ABS spectrum at $k_x = 0$.  For convenience, we rearrange the basis as $(c_{k_x=0,\uparrow},c_{k_x=0,\downarrow}, c_{-k_x=0,\downarrow}^{\dagger},- c_{-k_x=0,\uparrow}^{\dagger})^{T}$
	so that the $k_x = 0$ BdG Hamiltonian becomes
	\begin{align}
		H_{BdG} &= \left[\frac{\partial_y^2}{2m}-\mu(y)+\tilde h \sigma_x\right]\rho_z + i \tilde \alpha_R \partial_y \sigma_x \rho_z
		\nonumber \\&  + {\rm Re}\Delta(y) \rho_x + {\rm Im} \Delta(y) \rho_y,
	\end{align}
	where $\rho$ are  Pauli matrices that act in particle-hole space.
	The analysis is streamlined by the fact that, at $k_x = 0$, the BdG Hamiltonian in the preceding basis commutes with $\eta \equiv \sigma_x$.  
	In sector $\eta = \pm 1$, the Hamiltonian reduces to a $2\times2$ matrix
	\begin{equation}
		H_{\eta}=\begin{pmatrix}
			\xi(y)+i \eta \tilde \alpha_R \partial_y  +\eta \tilde{h}&\Delta(y) \\
			\Delta(y)^* &-\xi(y)-i \eta \tilde \alpha_R \partial_y +\eta \tilde{h}
		\end{pmatrix}
	\end{equation}
	with $\xi(y)=\partial_y^2/(2m)-\mu(y)$.  Next, we find the ABS energies using scattering matrix formalism.
	
	In the barrier region ($\Delta=0$, $\mu=\mu_1$), the energies are 
	$\epsilon=\pm [-k_y^2/(2m)-\mu_1]\pm \eta \tilde \alpha_R k_y+\eta \tilde{h}$ with associated wavefunctions 
	\begin{align}
		\psi_{e,\eta}^{\nu}=\begin{pmatrix}
			1\\0
		\end{pmatrix}e^{i k_{F,e}^\nu y}, \\
		\psi_{h,\eta}^{\nu}=\begin{pmatrix}
			0\\1
		\end{pmatrix}e^{i k_{F,h}^\nu y},
	\end{align}
	where
	\begin{align}
		k_{F,e}^{\nu} &= \eta m\tilde \alpha_R+\nu \sqrt{m^2\tilde \alpha_R^2+2m(\eta \tilde{h}-\epsilon-\mu_1)} \\
		k_{F,h}^{\nu} &= \eta m\tilde \alpha_R-\nu \sqrt{m^2\tilde \alpha_R^2-2m(\eta \tilde{h}-\epsilon+\mu_1)}  
	\end{align}
	%$k_{F,e}^{\nu}=\eta m\alpha+\nu \sqrt{m^2\alpha^2+2m(\eta \tilde{h}-\epsilon-\mu)}$, $k_{F,h}^{\nu}=\eta m\alpha-\nu \sqrt{m^2\alpha^2-2m(\eta \tilde{h}-\epsilon+\mu)}$ 
	and $\nu=\pm$ labels the right versus left movers.
	
	In the superconducting regions ($\Delta\neq0$, $\mu=\mu_2$), the wavefunctions are instead
	\begin{align}
		\psi_{e,\eta}^{'\nu} &= \frac{1}{\sqrt{2}}\begin{pmatrix}
			e^{i{\rm sgn}(y)\phi/2}\\e^{-i \gamma}
		\end{pmatrix}e^{ik_{S,e}^{\nu}y}\\
		\psi_{h,\eta}^{'\nu} &= \frac{1}{\sqrt{2}}\begin{pmatrix}
			e^{i{\rm sgn}(y)\phi/2}\\e^{i \gamma}
		\end{pmatrix}e^{ik_{S,h}^{\nu}y},
	\end{align}
	%
	% 	\begin{align}
	% 		\psi_{e,\eta}^{'\nu}=\frac{1}{\sqrt{2}}\begin{pmatrix}
	% 			e^{i\phi}\\e^{-i\text{arccos}\frac{\epsilon-\eta \tilde{h}}{\Delta}}
	% 		\end{pmatrix}e^{ik_{S,e}^{\nu}y}\\
	% 		\psi_{h,\eta}^{'\nu}=\frac{1}{\sqrt{2}}\begin{pmatrix}
	% 			e^{i\phi}\\e^{i\text{arccos}\frac{\epsilon-\eta \tilde{h}}{\Delta}}
	% 		\end{pmatrix}e^{ik_{S,h}^{\nu}y}
	% 	\end{align}
	%
	where  $\gamma=\text{arccos} \left( \frac{\epsilon-\eta \tilde{h}}{\Delta} \right)$,
	\begin{align}
		k^{\nu}_{S,e} &= k_{F,e}^{\nu}+\frac{i\nu\sqrt{\Delta^2-(\eta \tilde{h}-\epsilon)^2}}{\sqrt{m^2\tilde \alpha_R^2+2m(\eta \tilde{h}-\epsilon-\mu_2)}} \\
		k^{\nu}_{S,h} &= k_{F,h}^{\nu}+\frac{-i\nu\sqrt{\Delta^2-(\eta \tilde{h}-\epsilon)^2}}{\sqrt{m^2\tilde \alpha_R^2-2m(\eta \tilde{h}-\epsilon+\mu_2)}} .
	\end{align}
	%
	%$k^{\nu}_{S,e}=k_{F,e}^{\nu}+\frac{i\nu\sqrt{\Delta^2-(\eta \tilde{h}-\epsilon)^2}}{\sqrt{m^2\alpha^2+2m(\eta \tilde{h}-\epsilon-\mu)}}$,and $k^{\nu}_{S,h}=k_{F,h}^{\nu}+\frac{-i\nu\sqrt{\Delta^2-(\eta \tilde{h}-\epsilon)^2}}{\sqrt{m^2\alpha^2-2m(\eta \tilde{h}-\epsilon+\mu)}}$.
	
	With the above modes, it is straightforward to obtain the scattering matrix according to Refs.~\onlinecite{Beenakker1991, Ming2021}.  We first focus on the case with a uniform chemical potential in the junction and superconducting regions, i.e., $\mu_1=\mu_2 = \mu$. In the Andreev limit $|\mu| \gg (\Delta, B)$, the reflection matrices at the left interface $r_\text{L}$ and right interface $r_\text{R}$, and transmission matrices (left to right as $t_\text{RL}$ and right to left as $t_\text{LR}$) are
	\begin{equation}
		r_{\rm L}=\begin{pmatrix}
			0&e^{-i\gamma-i\frac{\phi}{2}}\\e^{-i\gamma+i\frac{\phi}{2}}&0
		\end{pmatrix}, 	r_{\rm R} =\begin{pmatrix}
			0&e^{-i\gamma+i\frac{\phi}{2}}\\e^{-i\gamma-i\frac{\phi}{2}}&0
		\end{pmatrix},
	\end{equation}
	
	\begin{equation}
		t_{\rm LR}=\begin{pmatrix}
			e^{ik_{F,e}^+ L}&0\\
			0&e^{ik_{F,h}^+ L}
		\end{pmatrix}, 	t_{\rm RL}=\begin{pmatrix}
			e^{-ik_{F,e}^- L}&0\\0&	e^{-ik_{F,h}^- L}
		\end{pmatrix}.
	\end{equation}
	The ABSs energies $\epsilon_n$  can be solved from $\text{Det}[1-r_{\rm L} t_{\rm LR} r_{\rm R} t_{\rm RL}]=0$, which  gives
	\begin{equation}
		\cos [2\beta_m L -2\gamma ]=\cos(\phi).
	\end{equation}
	Here, $\beta_m=\frac{k_{F,e}^+-k_{F,h}^-}{2}\approx \frac{\eta \tilde{h}-\epsilon}{v_F}$. The main text Eq.~\eqref{ABS_energy}  then follows by considering the short-junction limit where the Thouless energy satisfies $E_T=\frac{\pi v_F}{2L}\gg \Delta,h$.
	
	Now we examine the case with a chemical potential difference $\mu_1 \neq \mu_2$. The transmission matrices $t_{\rm LR}$ and $t_{\rm RL}$ keep the same form as above, though the reflection matrices now encode additional normal reflections. The matrix $r_{\rm L}$ can be written as
	\begin{equation}
		r_{\rm L} = \begin{pmatrix}
			ir e^{i\theta_0} &\sqrt{1-r^2} e^{i\theta_0}e^{-i\frac{\phi}{2}}\\
			\sqrt{1-r^2} e^{i\theta_0}e^{i\frac{\phi}{2}}& ire^{i\theta_0}
		\end{pmatrix},
	\end{equation}
	where
	\begin{align}
		r e^{i\theta_0}=\frac{2\sin \gamma (k_1^2-k_2^2)}{(k_1+k_2)^2e^{i\gamma}+(k_1^2-k_2^2)e^{-i\gamma}},\\
		\sqrt{1-r^2} e^{i\theta_0}=\frac{4k_1k_2}{(k_1+k_2)^2e^{i\gamma}+(k_1^2-k_2^2)e^{-i\gamma}},
	\end{align}
	and   $k_{1,2} \approx \sqrt{m^2\tilde \alpha_R^2+2m|\mu_{1,2}|}$.
	If $\mu_1=\mu_2$, one finds that $r=0$ and $\theta_0=-\gamma$, thus recovering the preceding results. The reflection matrix $r_{\rm R}$ can be obtained  by replacing $\phi \rightarrow -\phi$ in $r_{\rm L}$. One can readily verify unitarity of the reflection matrices $r_{\rm L}$ and $r_{\rm R}$. 
	
	\begin{figure}
		\centering
		\includegraphics[width=0.8\linewidth]{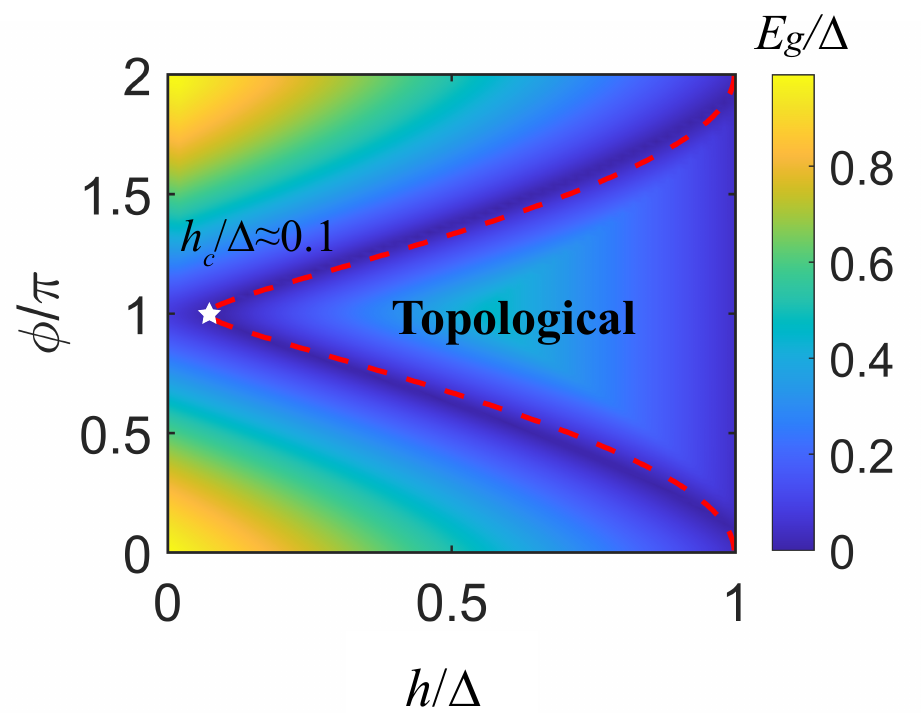}
		\caption{The $k_x = 0$ gap $E_g$ as a function of the Zeeman energy $h$ and phase difference $\phi$ calculated from the tight-binding model reviewed in Sec. V.  Parameters used are the same as in the main text Fig.~\ref{fig:fig3}(c), except that $\mu_1=3$ meV.  The red dashed lines are obtained from Eq.~(\ref{Eq_step}) with a normal reflection coefficient  $r=0.1$, while the white star indicates the critical field at $\phi=\pi$ extracted from the tight-binding model.}%the parameters used are the same as in the main text, Fig.~\ref{fig:fig3}(h), except for $\mu_1=3$ meV.}
		\label{fig:fig6}
	\end{figure}

	By solving  $\text{Det}[1-r_{\rm L} t_{\rm RL}r_{\rm R} t_{\rm LR}]=0$, we find that ABSs energies are determined by 
	\begin{equation}
		\cos (2\theta+2\beta_m L)+ r^2\cos(2\beta_p L)=(1- r^2)\cos(\phi). \label{Eq_normal}
	\end{equation}
	Here, the average Fermi wavelength between the left-moving and right-moving states reads $\beta_p=\frac{k_{F,e}^{+}-k_{F,e}^-}{2}\approx k_1$. In the short junction limit with weak normal reflections, we can approximate Eq.~\eqref{Eq_normal} as
	\begin{equation}
		\cos(2\gamma)=(1-r^2)\cos(\phi).\label{Eq_step}
	\end{equation}
	Near $\phi=\pi$, we can roughly obtain the topological phase transition line as $\tilde{h}/\Delta=|\cos(\phi/2)|+r/\sqrt{2}\sin(\phi/2)$.
	
	Let us compare the topological transition line given by Eq.~(\ref{Eq_step}) to our tight-binding model calculation (see Sec. V later), as we have done for the main text Fig.~\ref{fig:fig3}(c).  In  Fig.~\ref{fig:fig3}(c), the chemical potential difference between the superconducting and barrier regions does not produce obvious effects.  Here, we instead choose $\mu_1=3$ meV, $\mu_2=-1$ meV to amplify the role of chemical potential mismatch. In this case, the gap $E_g$ (extracted from the tight-binding model) as a function of Zeeman energy $h$ and phase difference $\phi$ is shown in Fig.~\ref{fig:fig6}. Note that the effective Zeeman energy felt by the partially occupied bands is weakly renormalized (i.e., $\tilde{h}\approx h$) 
	when $\lambda_0 \gg \beta_I$, as arises here. The red line is calculated from Eq.~(\ref{Eq_step}) with a normal reflection magnitude  $r=0.1$, which shows good agreement with the phase transition line indicated by $E_g$. The dominant change to the phase diagram occurs near $\phi=\pi$, where a finite Zeeman energy of $r\Delta/\sqrt{2}$ is now needed to drive the junction into the topological phase, contrary to the $r=0$ case where arbitrarily weak fields suffice.
	
	\section{Incorporating nematicity}
	
	\begin{figure}
		\centering
		\includegraphics[width=1\linewidth]{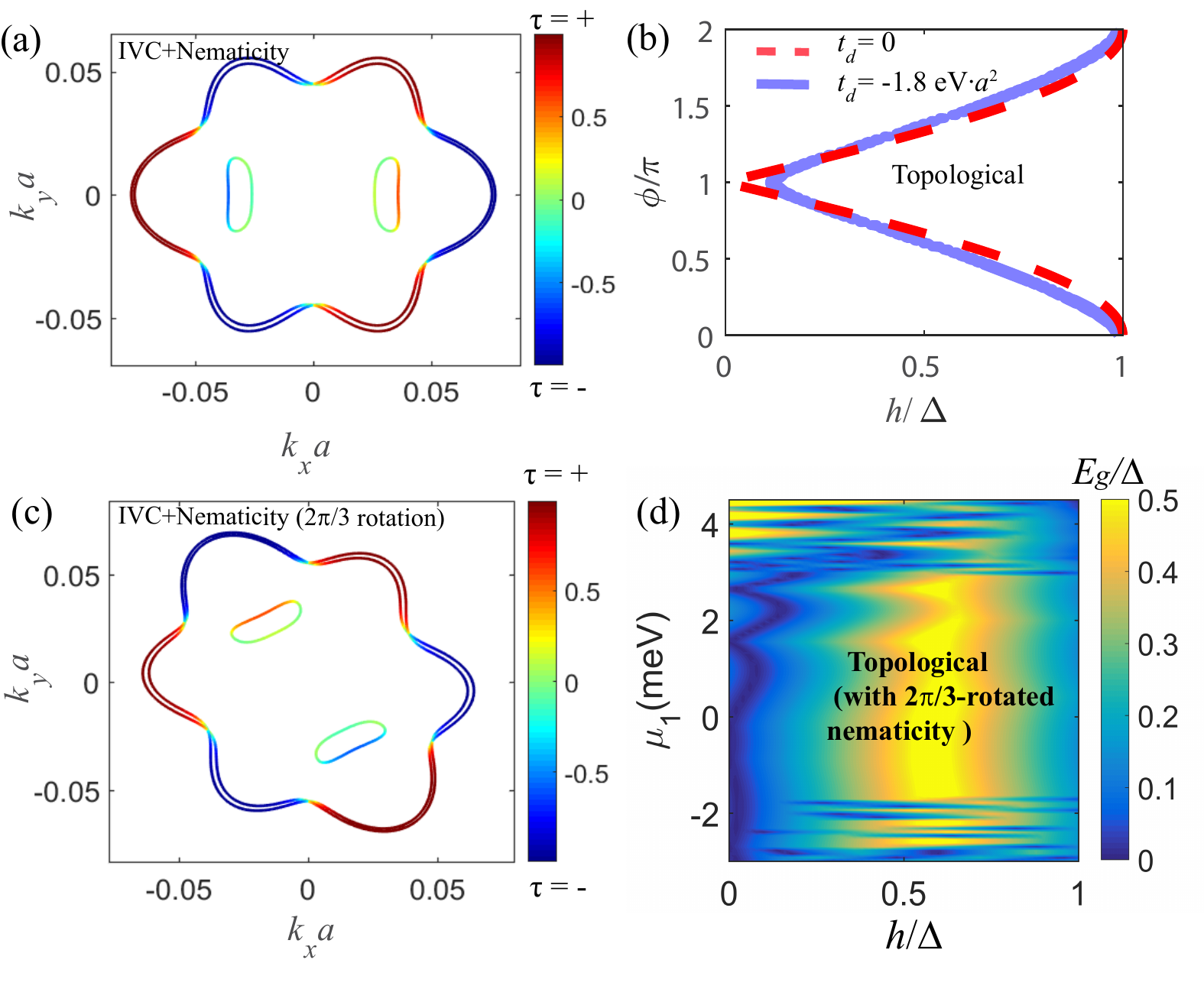}
		\caption{(a,c) Fermi contours in the presence of finite nematicity modeled with $t_d =-3$ eV $\cdot a^2$. Panel (c) differs from (a) by a $2\pi/3$ rotation on the nematicity direction. Other parameters are the same as for the main text Fig.~\ref{fig:fig2}(c). (b) Topological phase transition lines in the case with nematicity ($t_d=-1.8$ eV$\cdot a^2$) and without nematicity ($t_d=0$). Other parameters are the same as for the main text Fig.~\ref{fig:fig3}(f). Near $\phi=\pi$, the critical magnetic field to enter into the topological regions is modified slightly due to normal reflections enhanced by $t_d \neq 0$ in this case. (d)  Gap $E_g$ versus $h$ and $\mu_1$ obtained assuming the Fermi surface structure from (c).  Other parameters are the same as for the main text Fig.~3(d).  }
		\label{fig:fig9}
	\end{figure}
	
	Experiments \cite{zhang2023, Holleis2023} suggest that the number of small Fermi pockets present in the superconducting regime may be smaller than six due to nematicity. 
	To phenomenologically incorporate such effects, we explicitly break $C_3$ symmetry by replacing $\xi_0(\bm{k}) \rightarrow \xi_0(\bm{k})+t_d k_y^2$ in $h_0(\bm{k})$ from the main text. Figure~\ref{fig:fig9}(a) illustrates the resulting Fermi contours---which for chemical potential and $t_d$ value used exhibits only two small pockets.   In the range $t_d \sim -6$ to $-1.5$ eV$\cdot a^2$, the Fermi pockets look similar: large pockets plus two small pockets near $k_y=0$. A smaller $t_d$ would introduce other small pockets, while a larger $t_d$ would dramatically distort the large pockets.  We also verified that the chemical potential window at which the Fermi surface hosts only large Fermi pockets is only weakly affected by the introduction of $t_d$. 
	The influence of nematicity (modeled in this fashion) on the topological region in the $h$-$\phi$ and $h$-$\mu_1$ planes appear in Fig.~\ref{fig:fig9}(b) and the main text Fig.~3(e), respectively.  It can be seen that the nematicity does not appreciably affect the topological region in the $h$-$\phi$ plane. However, the robust topological region significantly broadens in the $h$-$\mu_1$ plane by removing the small pockets near $k_x=0$ through the presence of nematicity.
	
	In the preceding analysis we explicitly specified which two of the six small pockets remained in the presence of nematicity. 
	In principle, one can rotate the nematicity direction so that a different pair of two small pockets appears near Fermi energy. We find that the robust topological regime is mainly hampered by states near zero momentum along the junction (in our setup near $k_x=0$). In other words, provided the presence of nematicity can remove those states efficiently---which we find holds when the nematicity is strong enough to keep only two small pockets---the robust topological regime would be enlarged. Figure~\ref{fig:fig9}(c) illustrates a different set of two small pockets after a $2\pi/3$ rotation of the nematicity direction [$t_dk_y^2\mapsto t_d(\frac{\sqrt{3}}{2}k_x+\frac{1}{2} k_y)^2$] with respect to Fig.~\ref{fig:fig9}(a); as seen in Fig.~\ref{fig:fig9}(d), in this case a robust topological regime appears for $\mu_1\sim (-2,3)$ meV resembling that in the main text Fig.~3(e).

	\section{Effective tight-binding Hamiltonian for BLG/WSe2 Josephson junction}
	
	The tight-binding Hamiltonian used in the main text is deduced from the mean-field Hamiltonian Eq.~(\ref{Mean_field}) via a partial Fourier transform along $y$-direction.  As an illustration, we present the resulting tight-binding Hamiltonian where the $y$-direction (perpendicular to the junction) is open while $k_x$ is still a good quantum number. This tight-binding Hamiltonian can be written as 

	\begin{widetext}
		\begin{align}
			H_{tb} &=\sum_{j} \Big\{ \Psi^{\dagger}_{k_x,j} \left[ \left( (t_a+4t_c) k_x^2+t_ck_x^4-\mu_j+E_0 \right)
			+ \left(t_bk_x^3-6t_bk_x\right) \tau_z+\frac{\beta_I}{2}\tau_zs_z+\alpha_R k_xs_y+\lambda_0\tau_x+\lambda_1k_xs_y+hs_x \right] \Psi_{k_x,j}
			\nonumber \\
			&+\Psi^{\dagger}_{k_x,j} \left[ -\left(t_a+t_d+2t_ck_x^2+4t_c\right)+3t_b k_x \tau_z-\frac{\alpha_R}{2 i} s_x
			-\frac{\lambda_1}{2i}\tau_xs_x \right] \Psi_{k_x,j+1}+t_c\Psi^{\dagger}_{k_x,j}\Psi_{k_x,j+2} \Big\} \nonumber\\
			&+\sum_{j < -L/2}  \Delta e^{i\phi/2} \Psi^{\dagger}_{k_x,j} (i\tau_x s_y)  \Psi^{\dagger}_{-k_x,j} +  \sum_{j > L/2}  \Delta e^{-i\phi/2} \Psi^{\dagger}_{k_x,j} (i\tau_x s_y)  \Psi^{\dagger}_{-k_x,j} + \text{H.c.} 
			\label{eq:tight_binding}
		\end{align}
	\end{widetext}
	Here, $j$ labels lattice sites along the $y$ direction ($j=0$ corresponds to the middle of the junction); the four-component annihilation operator is $\Psi_{k_x,j}=(c_{+\uparrow,j}(k_x), c_{+\downarrow, j}(k_x),c_{-\uparrow, j}(k_x), c_{-\downarrow, j}(k_x))^{T}$; the chemical potential is $\mu_j=\mu_2 \theta(|j|a-L/2)+\mu_1 \theta(L/2-|j|a)$, and $E_0=2(t_a+t_d)+6t_c$ is a constant energy shift. We have defined the pairing matrix as $i\tau_x s_y$, which is an intervalley spin-singlet pairing. After projecting into the low-energy subspace spanned by Pauli matrices $\sigma_i$, this pairing takes the form $i\sigma_y$ as defined in the main text.
	
	A similar calculation using a tight-binding model to simulate the continuum model can be found in Refs.~\onlinecite{Manna2020, Ming2021}.
	Figures~\ref{fig:fig3} and \ref{fig:fig5} of the main text  and Figs.~(\ref{fig:fig6} to ~\ref{fig:fig_s6}) of the supplemental material are obtained from the above tight-binding model Hamiltonian $H_{tb}$. IVC order, when present, is always taken to be position-independent for simplicity, e.g., we assume that it is also present in the barrier of the gate-defined Josephson junction.  %\JA{Here is where we might comment that IVC order, when present, is always taken to be position-independent, i.e., we assume that it is present also in the junction.  This assumption seems like it might be benign if the ABS wavefunctions in the junction in practice have most of their weight in the superconducting regions.  In the short-junction limit that we invoke, is that scenario indeed borne out? And is this logic correct?  I'm not so sure... }\YM{We have set the IVC order parameter to be uniform in the space for simplicity. In practice, it can be different or even non-uniform between  inside and outside of the Josephson junction, depending on the interactions. I have specified this assumption more explicitly. }
	
	\section{The  minimal topological gap in various parameter regions}
	
	\begin{figure}[h]
		\centering
		\includegraphics[width=1\linewidth]{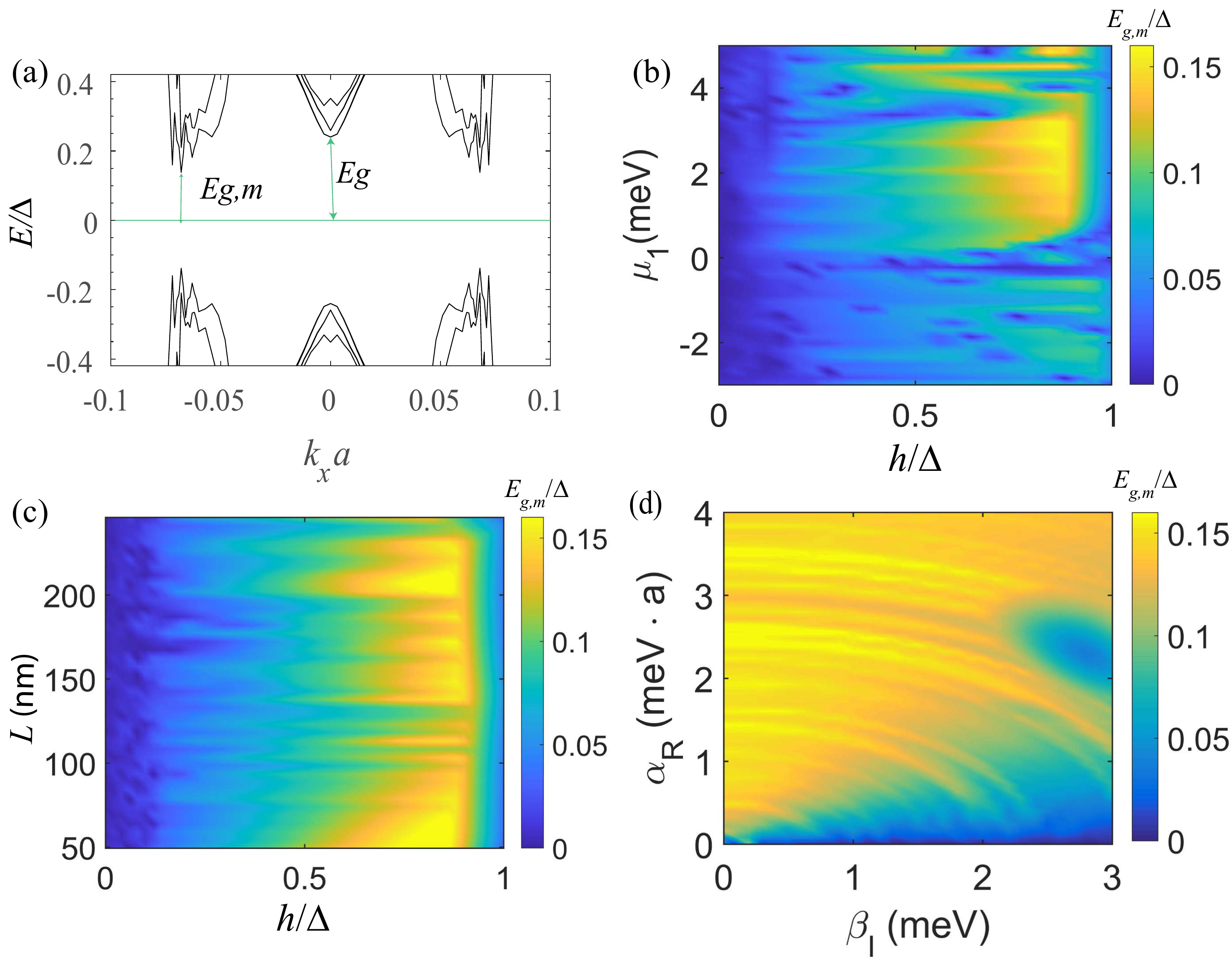}
		\caption{ (a) ABS spectrum versus $k_x$ obtained from the tight-binding model [Eq.~\eqref{eq:tight_binding}] with Zeeman field $h=0.8\Delta$. The gap $E_g$ at $k_x=0$ and the minimal gap $E_{g,m}$
			are highlighted. (b-d) Dependence of $E_{g,m}$ on various junction parameters.
			Data correspond to $L=78$ nm, $\mu_1=2$ meV, $\mu_2 = -1$ meV, $\beta_I=1.4$ meV, $\alpha_R=2 \text{meV}\cdot a$, $\lambda_0=3$ meV, $\lambda_1 = 0$,
			$h=0.8\Delta$, and $\phi=\pi$.}
		\label{fig:fig4}
	\end{figure}
	
	Figure~\ref{fig:fig4}(a) illustrates the $k_x$-dependent ABS spectrum for a topological phase with $\phi=\pi$ and $h=0.8\Delta$.  Notice that the $k_x = 0$ gap $E_g$ exceeds the minimal gap $E_{g,m}$.  The $E_{g,m}$ gap---which limits the decay length of MZMs in the barrier---typically arises at finite $k_x$ [Fig.~\ref{fig:fig4}(a)] and depends on various model parameters as shown in Fig.~\ref{fig:fig4}(b-d). 
	In our simulations the optimal $E_{g,m}$ approaches $\sim 0.2 \Delta$ and tends to occur $(i)$ over a broad range of lengths ($L \sim 50-200$ nm) in the short-junction regime;
	$(ii)$ for $\mu_1 \sim 0-3$ meV, where the barrier region is devoid of small Fermi pockets; and $(iii)$ when Rashba coupling energy $\alpha_R k_F$ is sufficiently large relative to Ising SOC. 
	Requirement $(iii)$ follows from the fact that Ising SOC renormalizes downward the effective Rashba SOC for the relevant large Fermi surfaces; recall Eq.~\eqref{parameters_projected} in the main text.  
	
	\section{Tunneling spectroscopy and Majorana zero-modes}
	
	\begin{figure}
		\centering
		\includegraphics[width=1\linewidth]{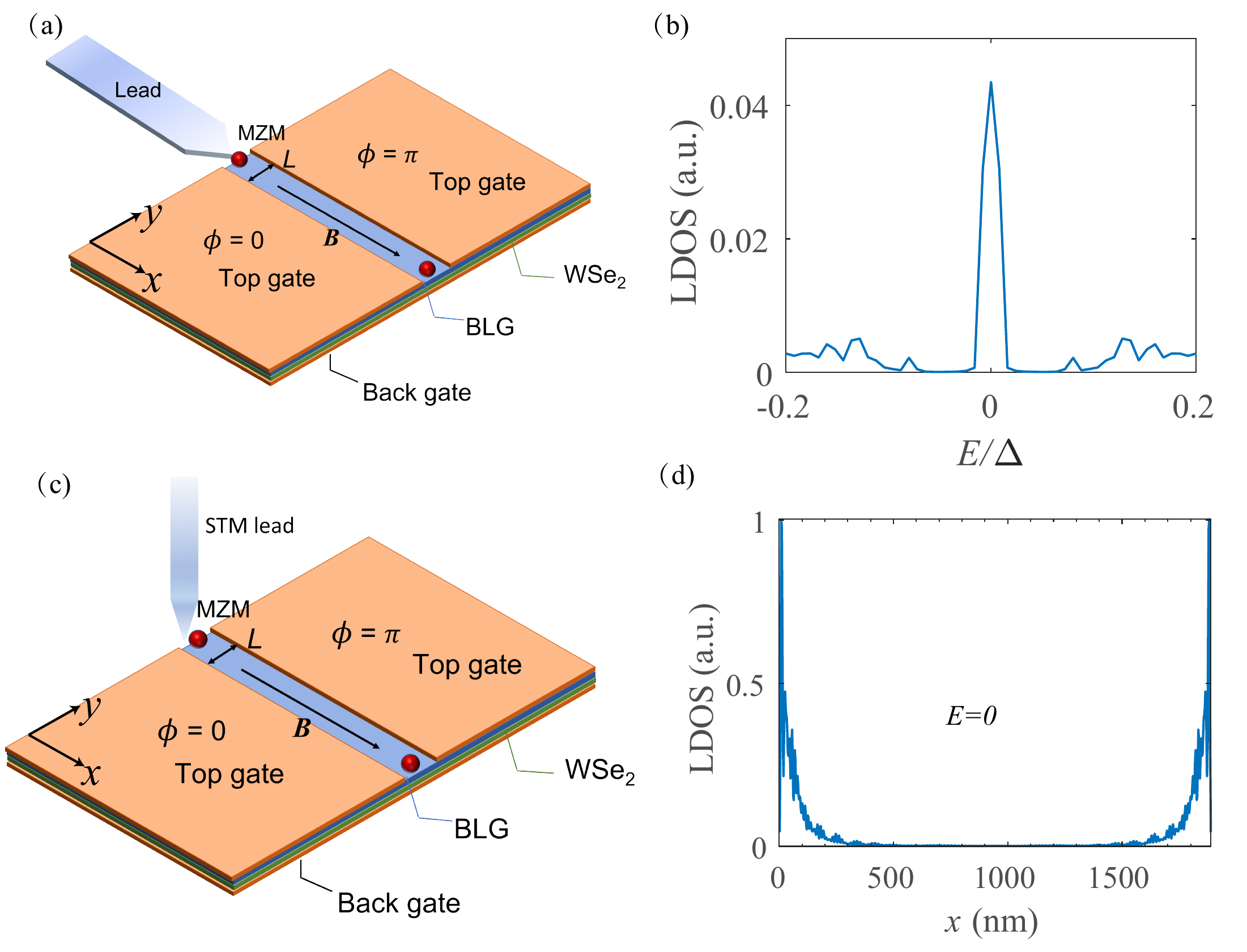}
		\caption{(a,c) Illustration of detection schemes for Majorana zero-energy modes (MZMs) in the BLG/WSe$_2$ planar Josephson junction.   (b,d) Local density of states (LDOS) that roughly mimics the conductance measured in the setups from (a,c).  Panel (b) shows the energy dependence of the LDOS near the end of the barrier in the topological phase of the Josephson junction. The zero-bias peak originates from the MZM localized at the boundary. 
			Panel (d) shows the position dependence of the zero-energy LDOS within the barrier---which maps the spatial structure of the Majorana modes.  Data in (b,d) were obtained assuming the following: $L=50$ nm, $\mu_1=2$ meV, $\mu_2 = -1$ meV, $\beta_I=1.4$ meV, $\alpha_R=2 \text{meV}\cdot a$, $\lambda_0=3$ meV, $\lambda_1 = 0$, $h=0.8\Delta$,$\phi=\pi$, and broading parameter $\eta=i\Delta/1000$.
		}
		\label{fig:fig_s5}
	\end{figure}
	
	The topological phase of the BLG/WSe$_2$ planar Josephson junction hosts a single MZM localized to each end of the barrier.  Tunneling constitutes a commonly deployed---though subtle to interpret unambiguously---experimental tool for diagnosing the presence of Majorana modes.  Figure~\ref{fig:fig_s5}(a,c) sketches two possible transport experiments to which our setup is amenable: (a) tunneling spectroscopy from a lead directly into the end of the barrier and (c) scanning tunneling microscopy (STM) that probes the full spatial extent of the barrier.  The lead in (a) could in principle arise from the BLG/WSe$_2$ medium itself upon introducing appropriate tunnel barriers via gates;
	while STM offers a broader spatial field of view, the accessible temperatures are comparatively high, possibly on the scale of or larger than the topological gap for the junction.  In this section we examine the LDOS in the barrier---which is expected to roughly track the conductances measured in such tunneling experiments. The LDOS at sites $\bm{r}_j$ is obtained from the Green's functions
	\begin{equation}
		\rho_{\rm LDOS}(\bm{r}_j)=-\frac{1}{\pi}\text{Im}\{\text{tr}\left[G(\bm{r}_j)\right]\}.
	\end{equation}
	Here, $G(\bm{r}_j)$ denotes the  Green's function at site $\bm{r}_j$, and the trace is taken over the valley and spin space. The LDOS is evaluated on a discrete grid of positions $\bm{r}_j$ used to define the effective tight-binding model Eq.~\ref{eq:tight_binding}---a discretized version of the low-energy continuum Hamiltonian in the main text. In the next section, we will show how to recover the full atomically resolved LDOS by introducing Wannier functions to describe atomic orbitals.
	
	We  obtain the LDOS from our tight-binding model [Eq.~\eqref{eq:tight_binding}] by Fourier also transforming along the $x$ direction, taking open boundary conditions to introduce endpoints for the barrier, and using the
	lattice Green's function method \cite{Manna2020, Ming2021} to obtain the Green's function $G_{nn}(E)=(E-H_{nn}-\Sigma+i\eta)^{-1}$ of the column representing the middle of the junction region, i.e., $y=0$, where $H_{nn}$ is the Hamiltonian of $n-$th column, $\Sigma$ is the self-energy due to the coupling between nearest columns, and $\eta$ is a broading parameter. See the caption of Fig.~\ref{fig:fig_s5} for parameters.  Figure~\ref{fig:fig_s5}(b) presents the energy dependence of the LDOS evaluated near the end of the barrier.  The pronounced zero-bias peak reflects the associated localized MZM, and is the counterpart of the (quantized) zero-bias peak that would arise at zero temperature in the transport setup from panel (a).   Figure~\ref{fig:fig_s5}(d) shows the spatial profile of the LDOS at zero energy, which is relevant for the STM setup from panel (b).  Well-localized MZM wavefunctions---one from each end---are clearly visible; with the (not unreasonable) parameters used here, the decay length is on the 100 nm scale.  
	\begin{figure*}
		\centering
		\includegraphics[width=1\linewidth]{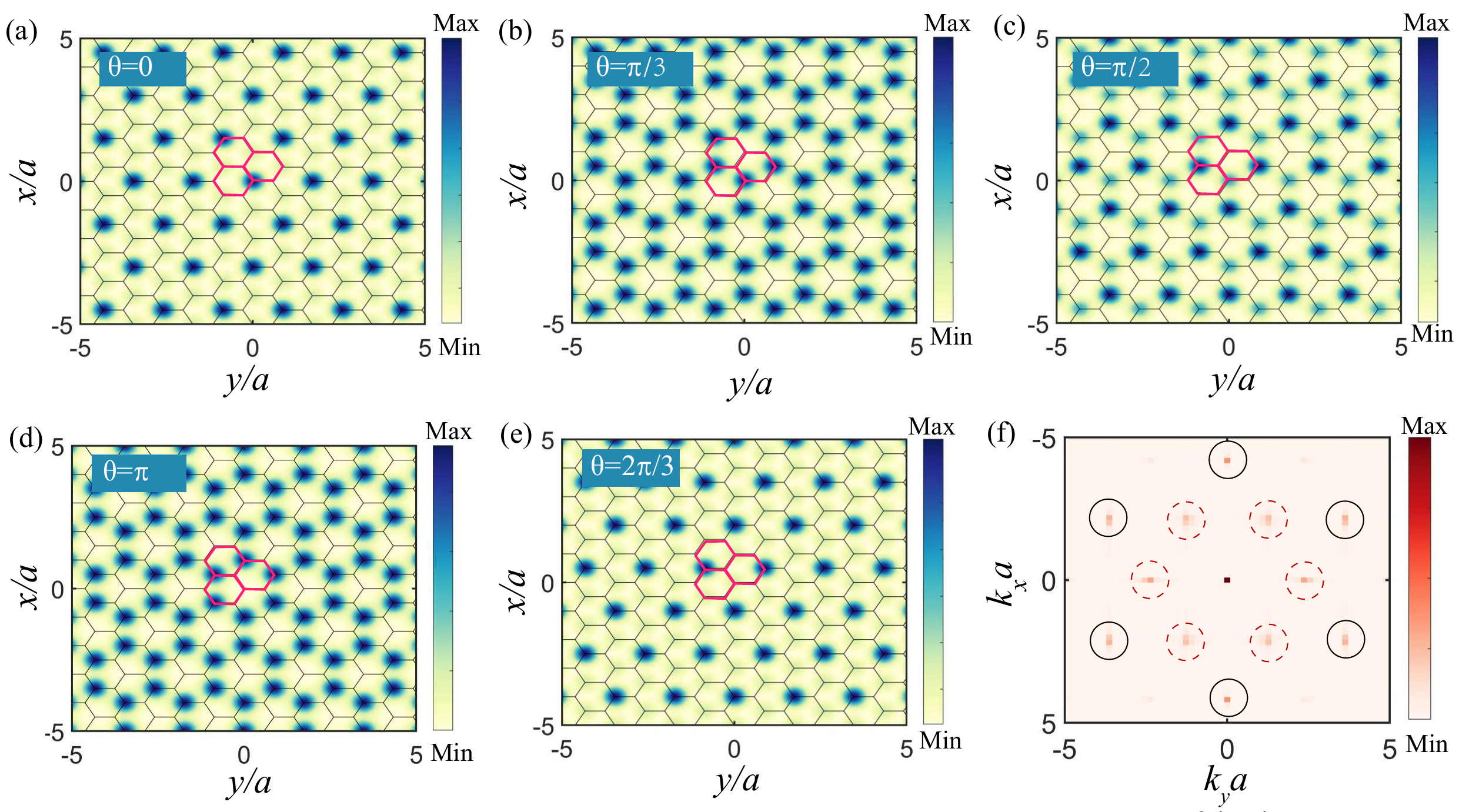}
		\caption{%Kekul\'e patterns corresponding to uniform IVC order in BLG. Panels 
			(a-e)  Kekulé patterns obtained from extended bulk states with energy $E =1.2\Delta$  at various Kekulé angles $\theta$. Three graphene hexagonal plaquettes are highlighted in pink color, which forms a  Kekulé supercell.  Note that mirror symmetry $M_x$ is preserved only for $\theta = 0$ and $\pi$ which correspond to a $\tau_x$ IVC order parameter (see SM Table~\ref{ref:table}). (f) Fourier transform of data in (a) for $\theta=0$ (other angles exhibit similar features). Kekulé peaks are highlighted with red dashed circles.}
		\label{fig:fig_s6}
	\end{figure*}
	\section{Computing atomically resolved Majorana zero mode wavefunctions}

	The formalism of the previous section allows us to evaluate a `coarse-grained' version of the LDOS, which captures the broad features of the low-energy wavefunctions but disregards atomically resolved information. We can capture the finer, atomic-scale structure of the LDOS by invoking Wannier functions that describe the relevant atomic orbitals. As shown in the main text, the necessity of IVC order to obtain topological superconductivity in our setup endows MZM wavefunctions with atomic-scale fingerprints, potentially detectable via STM.  In this section, we explain the method used to compute those atomically resolved MZM wavefunctions.
	
	For the purposes of this section, we parametrize the (spin-unpolarized) IVC order parameter as
	\begin{equation}
		\Delta_{\rm IVC} = \cos(\theta) \tau_x + \sin(\theta) \tau_y ,
		\label{eq:IVC_angle}
	\end{equation}
	where $\theta$ denotes the Kekulé angle.  Except for \text{Fig.~4} of the main text,  a $\tau_x$ order parameter (which respects the mirror symmetry $M_x$) was assumed, corresponding to $\theta=0$.  Different values of $\theta$ are contrasted in Fig.~\ref{fig:fig5} to aid with experimental identification of IVC orders. 
	
	The LDOS, now including atomic-scale structure, %\ELH{I am bit confused---shouldn't the LDOS always include atomic-scale structure? Kind of by definition. Same comment occurred in main text with somewhat clunky phrases like ``atomically resolved local density of states"}, 
	is given by
	\begin{equation}
		\rho(E,\bm{r})=-\frac{1}{\pi}\text{Im}\sum_{n}\frac{|\psi_{n}(\bm{r})|^2}{E-E_n+i\eta}
		\label{Eq_dos}
	\end{equation}
	where $\eta >0$ is a small broadening parameter and $\psi_{n}(\bm{r})$ is the wavefunction for an eigenstate with energy $E_n$ in the $n$-th band, 
	\begin{equation}
		\psi_{n}(\bm{r})=\sum_{\bm{r}_{\alpha},\tau,s}c_{n,\tau,s}(\bm{r}_{\alpha}) e^{i \bm{K}_{\tau}\cdot\bm{r}_{\alpha}} \phi_{\alpha}(\bm{r}-\bm{r}_{\alpha}).\label{Eigen_wave}
	\end{equation}
	Here $\tau$, $s$ respectively denote valley and spin indices, $\bm{K}_{\tau}=(\tau 4\pi/3a, 0)$, and $\phi_{\alpha}(\bm{r}-\bm{r}_{\alpha})$ represent localized Wannier wavefunctions at lattice site $\bm{r}_{\alpha}$. The coefficients $c_{n,\tau,s}$ can be obtained from diagonalizing the tight-binding Hamiltonian in Eq.~\eqref{eq:tight_binding} with the IVC order parameter now given by Eq.~\eqref{eq:IVC_angle}. For bulk states, $c_{n,\tau,s}$ are expected to be uniform in real space, while for MZMs in a junction geometry they display an exponential decay with localization length $\xi$. 
	
	Substituting Eq.~(\ref{Eigen_wave}) into Eq.~(\ref{Eq_dos}), the LDOS can be decomposed into two contributions: 
	$\rho(E,\bm{r})=	\rho_{0}(E,\bm{r})+\delta \rho(E,\bm{r})$. The first piece, 
	\begin{eqnarray}
		&&\rho_{0}(E,\bm{r})=-\frac{1}{\pi}\text{Im}\sum_{\tau\tau',\bm{r}_{\alpha}}\text{tr}[G_{\tau\tau'}(\bm{r}_{\alpha},\bm{r}_{\alpha})]\times\nonumber\\
		&&e^{i(\bm{K}_{\tau}-\bm{K}_{\tau'})\cdot \bm{r}_{\alpha}} |\phi(\bm{r}-\bm{r}_{\alpha})|^2,
	\end{eqnarray}
	involves Wannier orbitals at the same lattice site $\bm{r}_{\alpha}$, while the second, $\delta \rho(E,\bm{r})$, 
	includes all contributions with Wannier orbitals evaluated at different lattice sites.  
	In the equation above,  $G_{\tau\tau'}(\bm{r}_{\alpha}, \bm{r}_{\alpha}) $  is a Green's function %between site $\bm{r}_{\alpha}$ and $\bm{r'}_{\alpha}$ 
	(the spin indices are omitted for simplicity) that can be directly calculated from the tight-binding Hamiltonian, Eq.~\eqref{eq:tight_binding}, using the recursive Green's function method \cite{Manna2020, Ming2021}.  Since the relevant wavefunctions at the large $D$ fields of interest are well-localized on one layer, and on one sublattice, the contribution from $\delta \rho(E,\bm{r})$ is expected to be significantly suppressed compared to $\rho_0(E,\bm{r})$; moreover, $\rho_0(E,\bm{r})$ already suffices to capture atomic-scale structure descending from IVC order.  We therefore neglect $\delta \rho(E,\bm{r})$ hereafter.

	%To show how IVC states break the lattice translational symmetry on the graphene lattice scale, it is sufficient to consider  $\rho_{\rm sites}(E,\bm{r})$. But the bond density  $\rho_{\rm bonds}(E,\bm{r})$, which typically is smaller, can also be useful in revealing rotational symmetry breaking.
	
	To model the atomic  orbitals, we consider Gaussian Wannier functions~\cite{JungPyo2022,Hyunjin_2023}: 
	\begin{equation}
		\phi(\bm{r}-\bm{r}_{\alpha})=\frac{1}{\sqrt{2\pi}}e^{-\frac{(\bm{r}-\bm{r}_{\alpha})^2}{2\sigma^2}}.
	\end{equation}
	In our simulations, we take $\sigma = 0.3a$. Using this choice, the contribution $\delta \rho$ is indeed unimportant.   Due to the large displacement field $D$, we expect the valence-band wavefunctions will be pushed onto the $B_2$ sites (or the $A_1$ sites for the opposite sign of the displacement field). 
	Therefore, we position the Wannier centers of our tight-binding model on the $B_2$ sites, forming a triangular lattice.

	As an illustration, Fig.~\ref{fig:fig_s6}(a-e) plots the atomically resolved wavefunction of extended bulk states with energy $E =1.2\Delta$, for various Kekulé angles $\theta$. 
	%Figs.~\ref{fig:fig_s6}(a) to \ref{fig:fig_s6}(e) show the Kekulé pattern at various Kekulé angles $\theta$.  
	Unlike the Kekulé pattern for MZMs presented in the main text, here the patterns are translationally invariant---albeit with an enlarged unit cell reflecting IVC order. The main text Figs.~\ref{fig:fig5}(c-e) can thus be understood as 
	a uniform Kekulé pattern at the corresponding Kekulé angle, superimposed on the additional spatial modulations coming from the exponentially decaying spatial profile of MZMs. There is a small asymmetry between the LDOS at $y$ and $-y$ in the main text Figs.~\ref{fig:fig5}(c-e) arising from explicit breaking of mirror symmetry $M_y$ in the low-energy theory. This breaking is manifest in the second term of Eq.~(\ref{eff_Ham}), which arises from a combination of Ising SOC and trigonal warping. Figure~\ref{fig:fig_s6}(f) illustrates that, by similarly Fourier transforming the extended Kekulé bulk states, we obtain the usual Kekulé peaks in momentum space. 
\end{document}